\pdfoutput=1
\documentclass[nohyperref]{lecturenotes}

\def\HgNode{c2774373f504}

\def\HgRev{51}

\newif\ifsnapshot 
\snapshotfalse
\ifsnapshot
  \def\documentident{v1.x-SNAPSHOT \HgNode}
  \def\documentdate{\HgDate}
\else
  \def\documentident{v1.0}
  \def\documentdate{2012 August 17}
\fi

\title{DMP Planning for Big Science Projects}
\author{MaRDI-Gross project}
\date{\documentident, \documentdate}

\CopyrightStatement{Copyright 2012, University of Lancaster}

\usepackage{mardi-gross}

\ifsnapshot
\definecolor{draftgrey}{gray}{0.75}
\fi

\tocstyle{noleaders}

\ifxetex
  \PassOptionsToPackage{xetex,unicode}{hyperref}
\else
  \PassOptionsToPackage{pdftex}{hyperref}
\fi

\usepackage[hyperindex=false,pdftitle={DMP Planning for Big Science Projects},pdfauthor={Juan Bicarregui, Norman Gray, Rob Henderson, Roger Jones, Simon Lambert, Brian Matthews}]{hyperref}

\usepackage{glossaries}
\newglossaryentry{ATLAS}{name={ATLAS},description={A Large ToroidaL ApparatuS, physically the largest of the general purpose LHC detectors, and the associated collaboration that built it, operates it and exploits it}}
\newglossaryentry{CMS}{name={CMS},description={Compact Muon Solenoid, a general purpose LHC detector, and the associated collaboration that built it, operates it and exploits it}}
\newglossaryentry{LHC}{name={LHC},description={The Large Hadron Collider at \gls{CERN}: the accelerator is the host for two large general purpose detectors (ATLAS and CMS) and two smaller ones (ALICE and LHCb)}}
\newglossaryentry{LHCb}{name={LHCb},description={A dedicated LHC detector for the study of charm and beauty quarks, and the associated collaboration that built it, operates it and exploits it}}
\newglossaryentry{ESD}{name={ESD},description={Event Summary Data: the output data format from applying a High Energy Physics reconstruction programme to the RAW data},first={Event Summary Data (ESD)}}
\newglossaryentry{AOD}{name={AOD},description={Analysis Object Data: a reduced version of the \gls{ESD} intended for general purpose analysis},first={Analysis Object Data (AOD)}}
\newacronym{HEP}{HEP}{High Energy Physics}
\newglossaryentry{CERN}{name={CERN},description={European Centre for Particle Physics}}
\newglossaryentry{HERA}{name={HERA},description={A particle accelerator at the German DESY facility}}
\newglossaryentry{NSF}{name={NSF},description={National Science Foundation: the principal (non-defence) science funder in the USA},first={National Science Foundation (NSF)}}
\newglossaryentry{STFC}{name={STFC},description={the Science and Technology Facilities Council: the principal UK HEP, nuclear and astronomy funder (which in practice means \q{big science}, in the sense of international, multi-currency, collaborations); see \url{http://www.stfc.ac.uk}},first={the Science and Technology Facilities Council (STFC)}}
\newglossaryentry{RCUK}{name={RCUK},description={Research Councils UK: the \q{strategic partnership of the UK's seven Research Councils}},first={Research Councils UK (RCUK)}}
\newglossaryentry{ESRC}{name={ESRC},description={Economic and Social Research Council: the principal social science funder in the UK, see \url{http://www.esrc.ac.uk}},first={Economic and Social Research Council (ESRC)}}
\newglossaryentry{EPSRC}{name={EPSRC},description={Engineering and Physical Sciences Research Council: the UK funder for engineering, and all physics other than that covered by STFC \url{http://www.epsrc.ac.uk}},first={Engineering and Physical Sciences Research Council (EPSRC)}}
\newacronym{BBSRC}{BBSRC}{Biotechnology and Biological Sciences Research Council}
\newglossaryentry{NERC}{name={NERC},description={Natural Environment Research Council: the UK funder for research about the natural world \url{http://www.nerc.ac.uk}},first={Natural Environment Research Council (NERC)}}
\newglossaryentry{JISC}{name={JISC},description={Joint Information Systems Committee: The organisation responsible for the maintenance and effective exploitation of the academic computing network in the UK, and the funders of this present report},first={Joint Information Systems Committee (JISC)}}
\newglossaryentry{IVOA}{name={IVOA},description={International Virtual Observatory Alliance: the consortium which defines VO standards},first={International Virtual Observatory Alliance (IVOA)}}
\newglossaryentry{DASWG}{name={DASWG},description={Data Analysis and Software Working Group: the principal LIGO software forum},first={Data Analysis and Software Working Group (DASWG)}}
\newglossaryentry{VO}{name={VO},description={Virtual Observatory: a set of data sharing argreements and protocols, in particular the network of recommendations for sharing astronomy data, agreed through the IVOA (not to be confused with grid Virtual Organisations)},first={Virtual Observatory (VO)}}
\newglossaryentry{VOEvents}{name={VOEvents},description={The components of an alerting system for astronomical events, standardised by the IVOA}}
\newglossaryentry{strain data}{name={strain data},description={The fundamental GW signal}}
\newglossaryentry{data sharing}{name={data sharing},description={The formalised practice of making science data publicly available}}
\newglossaryentry{facility}{name={facility},description={A (typically large) nationally- or internationally-shared resource which scientists or groups will bid for time on; see \prettyref{s:bigscience}},plural={facilities}}
\newacronym{DMP}{DMP}{Data Management \& Preservation}
\newglossaryentry{DCC}{name={DCC},description={Digital Curation Centre: \url{http://www.dcc.ac.uk} (not to be confused with the LSC Document Control Center)},first={Digital Curation Centre (DCC)}}
\newglossaryentry{FITS}{name={FITS},description={Flexible Image Transport System: the standard file format in astronomy, see \url{http://fits.gsfc.nasa.gov}},first={Flexible Image Transport System (FITS)}}
\newglossaryentry{data products}{name={data products},description={Formal data outputs from an observatory, instrument or process}}
\newglossaryentry{pipeline}{name={pipeline},description={A software system (or sometimes a software-hardware hybrid) which transforms raw data into more or more levels of data product.  The data reduction pipelines, which must be able to keep up with the rate at which data is acquired, and which is assembled from a mixture of standard and custom software components, generally absorb a significant fraction of the total development budget of a new instrument}}
\newglossaryentry{raw data}{name={raw data},description={The data extracted directly from an instrument or observation; since it is uncalibrated and uncorrected, it is generally of little use to those not intimately familiar with the instrument}}
\newglossaryentry{proprietary period}{name={proprietary period},description={In the context of data release, a period extending for perhaps 12, 18 or 24 months after the data is taken, during which only the scientist who requested it can retrieve it, but after which it automatically becomes retrievable by anyone (\q{embargo} would be a better term, though unconventional)}}
\newglossaryentry{catalogue}{name={catalogue},description={In the astronomical context, a catalogue is a table of positions and other information for stars or other other astronomical objects}}
\newglossaryentry{KRDS}{name={KRDS},description={Keeping Research Data Safe: JISC project developing and documenting data preservation tools and studies; see \url{http://www.beagrie.com/krds.php} and \cite{beagrie10}},first={Keeping Research Data Safe (KRDS)}}
\newglossaryentry{TRAC}{name={TRAC},description={Trustworthy repositories audit and certification: a standard for accredition of archives \cite{dale07}},first={Trustworthy repositories audit and certification (TRAC)}}
\newacronym{PNM}{PNM}{Preservation Network Model}
\newglossaryentry{SKA}{name={SKA},description={Square Kilometre Array: a low frequency radio telescope with a large (one square kilometre) collecting area},first={Square Kilometre Array (SKA)}}
\newglossaryentry{LIGO}{name={LIGO},description={Laser Interferometer Gravitational-wave Observatory: the hardware, comprising LIGO Lab and GEO (see \url{http://ligo.org})}}
\newglossaryentry{aLIGO}{name={aLIGO},description={Advanced LIGO: The successor project to LIGO, due to start in 2015},first={Advanced LIGO (aLIGO)}}
\newglossaryentry{LIGO Lab}{name={LIGO Lab},description={The Caltech/MIT consortium, funded by NSF to design and run the LIGO interferometers in the US, see \url{http://www.ligo.caltech.edu}}}
\newglossaryentry{LSC}{name={LSC},description={LIGO Scientific Collaboration: The network of research groups contributing effort to the LIGO experiment and data analysis, see \url{http://ligo.org}},first={LIGO Scientific Collaboration (LSC)}}
\newglossaryentry{LVC}{name={LVC},description={A data-sharing agreement between the LSC and the Virgo Collaboration}}
\newglossaryentry{Virgo}{name={Virgo},description={Italian-French gravitational-wave detector \url{http://www.virgo.infn.it/}}}
\newglossaryentry{GEO}{name={GEO},description={A German-British consortium, responsible for the GEO600 interferometer, funded jointly by STFC and the German government}}
\newglossaryentry{GEO600}{name={GEO600},description={The GEO observatory located near Hannover in Germany}}
\newglossaryentry{RDS}{name={RDS},description={Reduced data sets: the core hierarchy of LIGO data sets},first={Reduced data sets (RDS)}}
\newacronym{GW}{GW}{Gravitational Wave}
\newglossaryentry{big science}{name={big science},description={A class of science projects characterised by being international, highly collaborative and expensively funded}}
\newglossaryentry{MRD}{name={MRD},description={Managing Research Data: a funding programme within the JISC e-Research theme, see \url{http://www.jisc.ac.uk/whatwedo/programmes/mrd}},first={Managing Research Data (MRD)}}
\newglossaryentry{CCSDS}{name={CCSDS},description={Consultative Committee for Space Data Systems: authors of the OAIS reference model, see \url{http://www.ccsds.org}},first={Consultative Committee for Space Data Systems (CCSDS)}}
\newglossaryentry{MOU}{name={MOU},description={Memorandum of Understanding: the relationships between the various participating entities in a collaboration is typically articulated through a series of MOUs, which may be fixed or periodically reviewed. These are not contracts, as such, but might cover reciprocal commitments of resources, and collaboration authorship policy},first={Memorandum of Understanding (MOU)},firstplural={Memoranda of Understanding (MOUs)}}
\newglossaryentry{CDS}{name={CDS},description={Strasbourg Data Centre: (see \url{http://cdsweb.u-strasbg.fr/})},first={Strasbourg Data Centre (CDS)}}
\newglossaryentry{ADS}{name={ADS},description={Astrophysical Data Service: a bibliographic archive for astronomy, based at the Harvard-Smithsonian Center for Astrophysics; ADS preserves full-text copies of journal articles, both in collaboration with publishers, and through a digitization process, and maintains a widely-used bibliographic ID system (\url{http://ads.harvard.edu})},first={Astrophysical Data Service (ADS)}}
\newglossaryentry{VizieR}{name={VizieR},description={A repository of astronomical catalogue data at \gls{CDS}}}
\newglossaryentry{arXiv}{name={arXiv},description={An electronic preprint service, see \url{http://arxiv.org}.  The arXiv started in the early 1990s, based on FTP and email.  It initially serviced particle physics and astronomy, but has expanded to cover other areas of physics, mathematics, and some areas of computing science}}
\newglossaryentry{NSSDC}{name={NSSDC},description={National Space Science Data Center: the permanent archive for NASA space science mission data \url{http://nssdc.gsfc.nasa.gov}},first={National Space Science Data Center (NSSDC)}}
\newglossaryentry{PDS}{name={PDS},description={Planetary Data System: the NASA data archive and standard set \url{http://pds.nasa.gov/}},first={Planetary Data System (PDS)}}
\newglossaryentry{NASA}{name={NASA},description={National Aeronautics and Space Administration: the US space agency \url{http://www.nasa.gov}},first={National Aeronautics and Space Administration (NASA)}}
\newglossaryentry{NARA}{name={NARA},description={National Archives and Records Administration: the US national archive \url{http://www.archives.gov}},first={National Archives and Records Administration (NARA)}}
\newglossaryentry{OCLC}{name={OCLC},description={\q{Founded in 1967, OCLC Online Computer Library Center is a nonprofit, membership, computer library service and research organization dedicated to the public purposes of furthering access to the world's information and reducing the rate of rise of library costs} \url{http://www.oclc.org}}}
\newglossaryentry{DOI}{name={DOI},description={Digital Object Identifier: \q{a system for identifying content objects in the digital environment. DOI\textregistered\ names are assigned to any entity for use on digital networks. They are used to provide current information, including where they (or information about them) can be found on the Internet. Information about a digital object may change over time, including where to find it, but its DOI name will not change.} \url{http://doi.org}},first={Digital Object Identifier (DOI)}}
\newglossaryentry{ESA}{name={ESA},description={European Space Agency: \url{http://www.esa.int}},first={European Space Agency (ESA)}}
\newglossaryentry{ESO}{name={ESO},description={European Southern Observatory: A pan-european agency running a set of southern-hemisphere telescopes \url{http://www.eso.org}},first={European Southern Observatory (ESO)}}
\newglossaryentry{OAI-PMH}{name={OAI-PMH},description={Open Archives Initiative Protocol for Metadata Harvesting: \q{The Open Archives Initiative Protocol for Metadata Harvesting (OAI-PMH) is a low-barrier mechanism for repository interoperability. Data Providers are repositories that expose structured metadata via OAI-PMH. Service Providers then make OAI-PMH service requests to harvest that metadata. OAI-PMH is a set of six verbs or services that are invoked within HTTP} \url{http://www.openarchives.org/pmh/}},first={Open Archives Initiative Protocol for Metadata Harvesting (OAI-PMH)}}
\newglossaryentry{tacit knowledge}{name={tacit knowledge},description={knowledge which remains in the heads of expert users rather than being explicitly documented; the experts may or may not know that they possess this knowledge, or that unexamined aspects of their practice are important (discussed vividly in~\cite{collins01} and extensively in for example~\cite{collins07})}}
\newglossaryentry{OAIS}{name={OAIS},description={Open Archival Information System: A standardised model of an archive; see \cite{std:oais}},first={Open Archival Information System (OAIS)}}
\newglossaryentry{Consumer}{name={Consumer},description={The role played by those persons, or client systems, who interact with OAIS services to find preserved information of interest and to access that information in detail. This can include other OAISs, as well as internal OAIS persons or systems (OAIS)}}
\newglossaryentry{Producer}{name={Producer},description={The role played by those persons, or client systems, who provide the information to be preserved. This can include other OAISs or internal OAIS persons or systems (OAIS)}}
\newglossaryentry{Data Object}{name={Data Object},description={Either a Physical Object or a Digital Object (OAIS) (that is, the `Data Object' is the sequence of bits, or the physical object which is \emph{the data} in the most primitive sense)}}
\newglossaryentry{Designated Community}{name={Designated Community},description={An identified group of potential Consumers who should be able to understand a particular set of information. The Designated Community may be composed of multiple user communities (OAIS)},plural={Designated Communities}}
\newglossaryentry{Digital Object}{name={Digital Object},description={An object composed of a set of bit sequences (OAIS)}}
\newglossaryentry{Information Object}{name={Information Object},description={A Data Object together with its Representation Information (OAIS)}}
\newglossaryentry{Representation Information}{name={Representation Information},description={The information that maps a Data Object into more meaningful concepts (OAIS)}}
\newglossaryentry{Representation Network}{name={Representation Network},description={The set of Representation Information that fully describes the meaning of a Data Object. Representation Information in digital forms needs additional Representation Information so its digital forms can be understood over the Long Term (OAIS).}}
\newglossaryentry{Retrieval Aid}{name={Retrieval Aid},description={An application that allows authorized users to retrieve the Content Information and PDI described by the Package Description.}}
\newglossaryentry{Content Information}{name={Content Information},description={The set of information that is the original target of preservation by the OAIS (OAIS)}}
\newglossaryentry{SIP}{name={SIP},description={Submission Information Package: An Information Package that is delivered by the Producer to the OAIS for use in the construction of one or more AIPs (OAIS).},first={Submission Information Package (SIP)}}
\newglossaryentry{AIP}{name={AIP},description={Archival Information Package: An Information Package, consisting of the Content Information and the associated {Preservation Description Information}, which is preserved within an OAIS (OAIS)},first={Archival Information Package (AIP)}}
\newglossaryentry{DIP}{name={DIP},description={Dissemination Information Package: The Information Package, derived from one or more AIPs, received by the Consumer in response to a request to the OAIS (OAIS).},first={Dissemination Information Package (DIP)}}
\newglossaryentry{PDI}{name={PDI},description={Preservation Description Information: The information which is necessary for adequate preservation of the Content Information and which can be categorized as Provenance, Reference, Fixity, and Context information (OAIS)},first={Preservation Description Information (PDI)}}
\newglossaryentry{Long Term}{name={Long Term},description={A period of time long enough for there to be concern about the impacts of changing technologies, including support for new media and data formats, and of a changing user community, on the information being held in a repository. This period extends into the indefinite future (OAIS)}}
\newglossaryentry{Information Package}{name={Information Package},description={The Content Information and associated Preservation Description Information which is needed to aid in the preservation of the Content Information. The Information Package has associated Packaging Information used to delimit and identify the Content Information and Preservation Description Information (OAIS)}}

\makeglossaries

\begin{document}

\maketitle

\noindent
\begingroup
\advance\columnwidth -2\fboxsep
\advance\columnwidth -2\fboxrule
\fbox{\begin{minipage}\columnwidth
\smallskip
\begin{list}{XXX}{%
    \leftmargin=2.5cm
    \listparindent=0pt
    \itemindent=0pt
    \labelwidth=2.5cm
    \labelsep=1em
    \itemsep=\smallskipamount
    \raggedright}
\item[Project] Managing Research Data Infrastructures for Big Science
(MaRDI-Gross, \url{http://purl.org/nxg/projects/mardi-gross})
\ifsnapshot
\else
  \item[Release] \documentident, \documentdate
  \item[URL] \url{http://purl.org/nxg/projects/mardi-gross/report}
\fi
\item[Authors]
Juan Bicarregui$^1$,
Norman Gray$^2$,
Rob Henderson$^3$,
Roger Jones$^3$,
Simon Lambert$^1$,
and 
Brian Matthews$^1$\\
$^1$E-Science Centre, STFC, UK;
$^2$School of Physics and Astronomy, University of Glasgow, UK;
$^3$Department of Physics, University of Lancaster, UK
\item[Distribution] Public
\end{list}
\smallskip
\end{minipage}
} 
\endgroup

\medskip

\section*{Abstract}

This report exists to provide high-level guidance for the strategic
and engineering development of \gls{DMP} plans \glsreset{DMP}
for \q{Big Science} data.

Although the report's nominal audience is therefore rather narrow, we
intend the document to be of use to other planners and data architects
who wish to implement good practice in this area.  For the purposes of
this report, we presume that the reader is broadly persuaded (by
external fiat if nothing else) of the need to preserve research data
appropriately, and that they have both sophisticated technical support
and the budget to support developments.

The goal of the document is not to provide mechanically applicable
recipes, but to allow the user to develop and lead a high-level plan
which is appropriate to their organisation.  Throughout, the report is
informed where appropriate by the OAIS reference model.

This report was funded by JISC in 2011--2012,
as part of the RDMP strand of the JISC programme
\href{http://www.jisc.ac.uk/whatwedo/programmes/mrd}{Managing Research Data}.
For document history, see p.\pageref{s:documenthistory}.

\newpage

\thispagestyle{empty}
\begingroup
\parindent=0pt
\parskip=\medskipamount

\null
\vfill
\includegraphics[height=10mm]{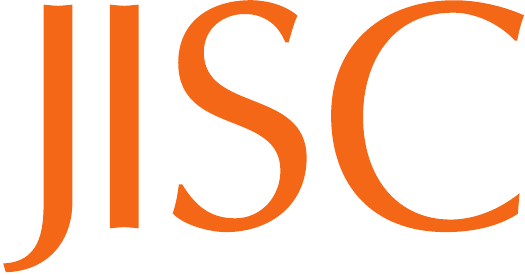}\\[\bigskipamount]
This report was prepared for, and funded by, the RDMP strand of the JISC programme
\href{http://www.jisc.ac.uk/whatwedo/programmes/mrd}{Managing Research Data}.

\includegraphics[scale=0.5]{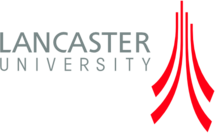}\\[0pt]
\includegraphics[height=10mm]{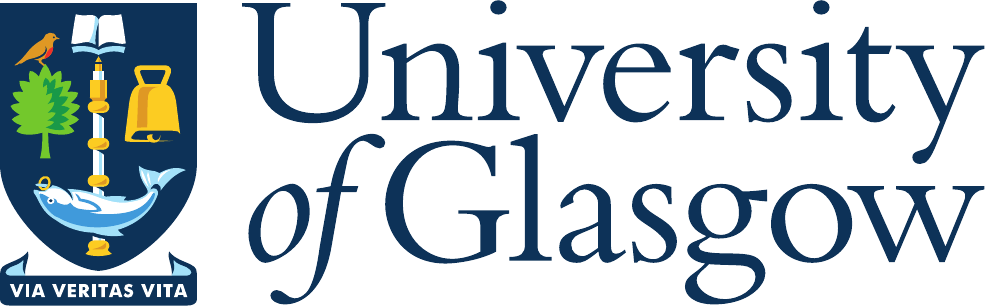}\\[\bigskipamount]
\includegraphics[height=10mm]{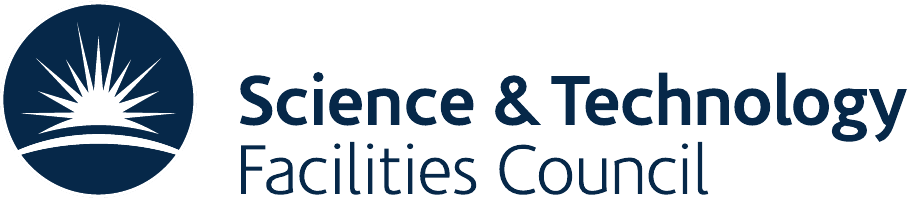}\\[\bigskipamount]
The MaRDI-Gross project was a collaboration between
the Department of Physics, University of Lancaster, UK;
the School of Physics and Astronomy, University of Glasgow, UK;
and the E-Science Center, STFC, UK.

Release: \documentident, \documentdate\ (\HgNode)

\makeatletter
\@CopyrightStatement.
\makeatother
This work is licensed under the Creative Commons
   Attribution 2.0 UK: England and Wales
   Licence. To view a copy of this licence, see
   \url{http://creativecommons.org/licenses/by/2.0/uk/}.
\endgroup

\newpage

\setcounter{tocdepth}2
\tableofcontents

\clearpage

\setcounter{section}{-1}        

\section{Introduction}

This document has a very specific audience.
It is addressed to people who have, or who have been landed
with, the responsibility for developing a \gls{DMP} policy for a
\q{big science} collaboration, or some similar
multi-institutional or multi-national project with a need for a bespoke plan.

Although it is nominally addressed to this (rather small) readership, we have written it with the intention that it will additionally be of use to:
\begin{itemize}
\item those evaluating or assessing such plans, for example within funders; and
\item people developing similar bespoke plans for scientific and other
  entities at this or other scales, who are looking for practical guidance on where to start,
  but for whom existing \gls{DMP} guidance is too low-level or mechanical.
\end{itemize}
For the purposes of this report, we presume that the reader is broadly
persuaded (by external fiat if nothing else) of the need to preserve
research data appropriately, and that they have both sophisticated
technical support and the budget to support bespoke developments where
necessary, obtained from a broadly supportive funder.  We take the
position that:
\begin{itemize}
\item the demand for principled data management and data sharing is a reasonable one,
  and note that publicly funded projects typically have no fundamental
  objections to it;
\item that a reasonable framework for at least approaching the problem
already exists in \gls{OAIS} (\prettyref{s:whyoais});
\item that the OAIS recommendation is (just) concrete
enough that it is not merely waffle; and
\item that there is a bounded set of resources which, if mastered by
  the reader, will allow them to produce a project DMP plan which is
  practically acceptable to the project, and discharges the principled
  demands of the funder and of society.
\end{itemize}
Within this report we have sought to represent a consensus of views 
across the \q{large-science} community within the UK, 
both through the roles of the authors of this document and also through a wider 
consultation we have undertaken with funders and research leaders.
For more specific acknowlegements, see the section on~p.\pageref{s:acknowledgements}.


The document is structured into three parts.
\begin{itemize}
\item \prettyref{s:policy}, policy background: this part discusses the various
  high-level policy drivers for DMP planning. We take it as read that
  an organisation is aware of the need to manage its data
  professionally, in order that this data is readily accessible to the
  researchers within it. However, there are a number of higher-level
  interests which must be respected, concerning longer-term
  disciplinary goals, and the goals of society at large.  
\item \prettyref{s:technical}, technical background: this part is mostly about the
  technical frameworks relevant to the good management of data, and in
  particular the \gls{OAIS} model.  We believe this is the key set of
  technologies which someone producing a project DMP policy should be
  aware of.
\item \prettyref{s:planning}, DMP planning: everything more specific, which includes
  some discussion of the (poorly-modelled) costs of such preservation, and
  of existing work on validating (and its conjugate, auditing) DMP
  plans.  Though this section is more detailed than the earlier ones,
  it is not concerned with the nitty-gritty of RAID, network or NAS
  management, which are the province of the DMP plan's implementers.
\end{itemize}

\q{Data management} does not contain many profound imponderables; navels
need not be gazed at. Though it is going too far to say that we are
peddling organised common sense, the majority of the relevant
background material is readily accessible, as long as it can be found,
and be known to be relevant. Our practical goal in this document is to
assemble and contextualise this background material, arrange it in a
way which is useful to the consituency we are aiming at, indicate
where best practice may be found or where it is still unknown,
and thereby enable the reader to
lead the development of a DMP plan for their organisation, secure in
the knowledge that they have a reasonable claim to be on top of the
relevant literature.

\subsection{Focuses, coverage, and some definitions}

The document is practical in tone, necessarily without
being prescriptive; however, for our intended audience, the \q{practical}
includes some aspects of the larger policy background which must be
respected, so we include coverage of these aspects, as well.

The report has been produced with a UK focus, but the only places where
this is, we believe, apparent are in the UK emphasis of the policy
discussion in \prettyref{s:rcuk-principles}, and on the prominence of
the \gls{STFC} in our definition of big science below.
Although STFC is (for this reason) particularly prominent, there is
\q{big science} data also to be found in research supported by the
\gls{EPSRC}, the \gls{BBSRC} and the \gls{NERC}.

There is more context available in the document \q{Managing Research
  Data in Big Science}~\cite{gray12}.  This is the final report of a
project funded by JISC in 2010\range 11, which was concerned with the
background for big-science data management in general, and this
present report in some places draws text directly from the earlier
one. This might be useful for fuller discussion or further references,
and we will make occasional reference to it in order to keep this
present document short.

Throughout, the report is informed where appropriate by the OAIS
reference model.  The model is introduced as technical background in
\prettyref{s:oais-description}, and more details are discussed in that
section and as details of practice in
\prettyref{s:practice-validation}, but the ideas are pervasive enough
that we feel it is useful to give a brief informal description of the
model and its advantages at the beginning of the document, in
\prettyref{s:whyoais}.

For clarity, it seems useful to make briefly explicit what we mean by
\gls{DMP} and the term \q{big science}, and we do this in the
subsections below.

\subsection{The what, why and how of OAIS}
\label{s:whyoais}

As suggested above, this document's advice orbits around the
\gls{OAIS} standard, adopting its (useful) concepts and vocabulary,
and making reference to the other work on validation and costing that
builds on it.  It is therefore useful to briefly discuss the \q{what?},
\q{why?} and \q{how?} of OAIS, in that order.

\emph{What is the OAIS model?}
The OAIS reference model~\cite{std:oais} is a conceptual model of the
functions and responsibilities of an archive of (typically) digital objects,
where the archive is viewed as an organisation or other entity, in principle distinct from
the data producer, which exists to preserve those objects into the
\gls{Long Term}.  The OAIS standard does not describe how to achieve
this, but it \emph{does} clearly articulate the various steps of the
process (for example that data goes through phases of Submission
to an archive, Preservation there, and Dissemination to
users), the various roles involved (for example data \glspl{Producer}
versus \glspl{Consumer}), and what, at a high level, has to be done to
let all this happen (for example the creation and management of
documentation about \gls{Representation Information}).  There is a
fuller description of OAIS in \prettyref{s:oais-description}.

\emph{Why should you care?}  Integral to its development, the OAIS
standard defines a fairly extensive vocabulary for digital
preservation (each of the capitalised terms in the preceding paragraph has
a precisely defined meaning, and when such terms appear below they are
included in the glossary at the end), and although none of these
definitions is particularly startling, and although the standard text
can seem a little verbose, verging on windy, these terms have become
the standard ones, and most work in this area is framed, directly or
indirectly, by the OAIS concept set.  Thus, although the OAIS model is
not the \emph{only} model for a digital archive (see
\prettyref{s:dcc-lifecycle} for another), it is both plausible and
conventional, and so makes a good starting point, and a useful shared
understanding, for any discussion of digital preservation.  In
addition, it is worth pointing out that the model was developed by the
\gls{CCSDS}, and so has a heritage which makes it a natural
fit for non-space science data.

\emph{How do I implement an OAIS model?}
There is no general recipe, and by assumption the readers of this
document are interested in systems which are large or unusual enough
that no recipe is likely to be applicable.  Instead, we aim to provide
pointers to resources which guide you in the right direction, and
possibly reassure you that there are no major areas of concern you have missed.
To start with, there is
the brief introduction below in \prettyref{s:oais-description},
plus tutorial reports such as~\cite{lavoie04},
and book-length resources such as~\cite{giaretta11}.

\emph{OK, how do I know \emph{when I have implemented} an OAIS model?}
The OAIS model can be criticised for being so high-level that
\qq{almost any system capable of storing and retrieving data can make
  a plausible case that it satisfies the OAIS conformance
  requirements}~\cite{rosenthal05}, so it is important to be able to
reassure yourself, as a data manager, that you have achieved more than
simply producing the statement \qq{we promise not to lose the data}, dressed
in OAIS finery.  This is the domain of \emph{OAIS certification}, and
this involves both efforts to define more detailed
requirements~\cite{rosenthal05}, and efforts to devise more stringent
and more auditable assessments of an OAIS's actual ability to be
appropriately responsive to technology
change (see~\cite{caspar-evaluation09} and \cite[ch.25]{giaretta11}, and \prettyref{s:caspar}).  The conjugate of validation is the
question of how, as a funder, you reassure yourself that the DMP plan
which a project has proposed is actually capable of doing what you
(and you hope the project) wish it to do.  Together, these are the
domain of \emph{OAIS auditing}, and this is discussed in
\prettyref{s:oais-audit}

\subsection{What is \q{big science}?}
\label{s:bigscience}
Big science projects tend to share many features which distinguish
them from the way that experimental science has worked in the
past.  The differences include big money, big author lists and, most
famously, big data: the \gls{aLIGO} project (for example) will produce
of order \SI1\PBY, comparable to the \gls{ATLAS} detector's \SI{10}\PBY; the
eventual SKA data volumes will dwarf these.  See \cite[\S1]{gray10}
for extended discussion of the characteristic features of large-scale
science.

While the large data volumes bring obvious complications,
there are other features of big science which change the way we can
approach its data management, and which in fact make the problem
easier.
\begin{itemize}
\item Big science projects are often well-resourced, with plenty of
  relevant and innovative IT experience, engineering management and
  clear collaboration infrastructure articlulated through
  \glspl{MOU}. This means that such projects
  can develop custom technical designs and implementations, to an
  extent that would be infeasible for other disciplines.
\item These areas have a long necessary tradition of using shared
  \glspl{facility}, so engineering discipline, documented interfaces and
  SLAs are familiar to the community.
\item Historical experience of \q{large} data volumes mean everyone
  knows that ad hoc solutions don't work. Part of the challenge of
  developing and deploying principled DMP plans in other disciplines
  is the challenge of persuading funders and senior project members
  that effective data management is necessary, expensive and technically
  demanding, and cannot be simply left to junior researchers, however
  \q{IT-literate} they may seem to be. This battle is won in disciplines
  with long experience of large-scale data.
\end{itemize}

In particular, the projects we are focusing on in this project\dash
and what we take the term \q{big science} to refer to in this
report\dash are the \q{facilities} and international projects
typically funded in the UK by \gls{STFC}.  A \q{facility} in this
context refers to a (typically large) resource, funded and shared
nationally or internationally, which scientists or groups will bid for
time on.\footnote{A telescope's call for proposals is closely
  analogous to a grant funding call, except that the award will be
  nights in a forthcoming semester, rather than money.}  The facility
will be to some extent a \q{general purpose} device, such as a
telescope or an accelerator like ISIS.  Facilities represent major
infrastructural investments, typically enjoy a certain autonomy, and
are designed and managed through SLAs.
Facilities are generally highly automated, and typically take data
directly from the instrument into an archive.  This last point has multiple
implications for \gls{DMP}.

The \gls{LHC} and \gls{LIGO}
are probably too closely associated with particular goals and
collaborations to be naturally termed \q{facilities}, but they are of
the type of international project with the same data challenges.

Our definition of \q{big science} in this report is, to a first approximation,
roughly equivalent to \q{STFC-funded science}.  STFC is the UK's
primary big-science funding council, as it is structured with a
particular emphasis on multi-partner collaborative science, less
support than the other councils for few-person projects, and budgetary
arrangements with the UK Treasury which reflect its exposure to
long-term commitments in multiple currencies.  Although most STFC
science is \q{big science} by our definition, the converse is not
true, there are examples of such projects funded by both \gls{EPSRC}
and \gls{NERC}, and we hope that this document will be of use to
people in these areas, too.

The most obviously relevant feature of \q{big science} in our
definition is, of course, the \q{big data}
aspect.  Though not a defining feature, it is characteristic of such
projects that they are generally willing to deal with data volumes at
the upper end of what is feasible, if necessary by designing
instruments to produce data volumes no larger than what is predicted
to be manageable by the time the instrument finally comes on-line.  Without
discounting the technical achievements required by such data rates,
the key implication here is that day to day data \emph{management} is a core
concern of the project, which is designed and funded accordingly.
There are two key consequences of this, both positive.
\begin{itemize}
\item Data \emph{preservation}\dash meaning the continuance of the
  successfully managed data into the future\dash is straightforwardly
  identified as a cousin of the data \emph{management} problem.  The
  former problem is not trivial (in a sense expanded on in
  \prettyref{s:what-dmp}), since it has distinct goals and, for 
  example, a different budget profile, but some of the more
  troublesome aspects of \textit{ab initio} data preservation are
  handled for free by the necessary existence of a data management
  infrastructure.
\item In particular, the problems of data ingest, which loom so large
  in much of the \gls{DMP} literature, are reduced to the problem of
  documenting and possibly adjusting archival metadata.
\end{itemize}

Part of the motivation for this present document is the contention
that, for technologically sophisticated areas such as this one, the
guidance towards the development of a DMP plan can be boiled down to
\qq{Here's a copy of the \gls{OAIS} spec; get on with it}. 

\subsection{What is \q{data management and preservation}?}
\label{s:what-dmp}
The OAIS specification makes the general remark that
\qq{[t]ransactions among all types of organizations are being conducted
  using digital forms that are taking the place of more traditional
  media such as paper.  Preserving information in digital forms is
  much more difficult than preserving information in forms such as
  paper and film. This is not only a problem for traditional archives,
  but also for many organizations that have never thought of
  themselves as performing an archival
  function}~\cite[\S1.3]{std:oais}.

In the scientific context, \q{data management} has a somewhat narrower
remit: essentially all new scientific data, and a lot of scientific
metadata, is \q{born digital}, and is also born complete, in the sense
(expanded in~\cite[\S1.7]{gray10}) that the information to be archived
is designed and documented in such as way as to support future
scientific analysis.  Also\dash and this is common to most \glspl{facility}
science, and the envy of other disciplines\dash most large-scale
science data is acquired and archived automatically, in a system which
must be functioning adequately if the project as a whole is to
function at all, so that the matter of data \emph{preservation} at
first appears to be simply a question of copying data from a
day-to-day management system into a persistent archive.

But this is not the case.  In large and complicated experiments, 
the complication of the apparatus makes
it hard to communicate into the future a level of understanding
sufficient to make plausible use of the data.  This is discussed
below, in \prettyref{s:case-data-preservation}.

This is a useful place to stress that the OAIS definition of the
\gls{Long Term} is simple and pragmatic: the Long Term is, in effect, longer
than one technology generation, and thus far enough into the future that the
data will have to undergo some storage migration, and that future
users will have to depend on documentation rather than human contact
with the data creators.

This in turn leads naturally to the observation that data management
covers both storage -- the preservation of the bits -- and
curation -- the preservation of the knowledge about the bits.  The
storage problem is a technical and financial one: we will largely
avoid the technical question of which storage technology should be
used, save to note that answering this is part of the implementation
phase of a DMP plan and that the question must be re-answered by the
archive with each technology generation (we discuss storage technology
questions very briefly in \prettyref{s:storage}).  The financial aspect
to the storage problem is the question of how much it will cost to
store the data into the indefinite future: while storage costs for the
few-year short term can be trivially assessed with a couple of hours'
work on eBay, the unpredictability of the current long-running
decrease in storage prices means that long-term cost estimates are
both vital, if a solution is to be sustainable, and very poorly
understood.  For a discussion of the estimation of storage costs, see
\prettyref{s:preservation-costs}.

Curation costs, by contrast, are dominated by the front-loaded staff
costs for creating Representation Information documentation, and by
the non-negligible but broadly predictable staff costs of continuing
archive management.


\section{Policy -- the \q{why} of DMP planning}
\label{s:policy}

This part contains material about the larger-scale, \q{softer}, policy
context. The practical motivation for its inclusion here is that it
can provide the rationale for some of the aspirations and
prescriptions in the more concrete parts later.

\subsection{RCUK data principles and their interpretation}
\label{s:rcuk-principles}
In 2011, \gls{RCUK} developed and published a set of \q{Common Principles on
Data Policy}, intended to provide a framework for individual Research
Council policies~\cite{rcuk-principles}.   The RCUK principles
are informed by the earlier OECD
\q{Principles and Guidelines for Access to Research Data from Public
  Funding}~\cite{oecd-data-policy},
and in turn inform the discipline-specific policies of the various UK
research councils. 

In this section we compare the \gls{RCUK} and \gls{STFC} principles,
which are the ones of most immediate relevance to the big-science
disciplines of our study.  The aim is to give
texture to the otherwise rather sheer surfaces of the two sets of
principles, to make links between them and other sections of this
document, where appropriate, and to host some other remarks which do
not fit naturally anywhere else.


These are not, of course, the only sets of data sharing principles.
The \gls{EPSRC} requirements~\cite{epsrc-data-policy} are formulated
as a set of principles which are almost identical to the RCUK ones,
plus a set of \q{expectations} of features that will be present in the
final products of EPSRC-funded research, whilst avoiding being
restrictively specific about exactly how these expectations will be
satisfied.  The US's \gls{NSF} makes similarly generic demands, at the other
end of the funding process, that project submissions include a data
management plan with certain features~\cite{nsf11}.  The \gls{BBSRC}
\q{Data Sharing Policy}~\cite{bbsrc-data-policy} is somewhat more
specific, reflecting not just the different science, but the different
scale of science and the different technical expertise
available. Finally, the \gls{JISC}'s \q{Managing Research Data}
programme~\cite{jisc-mrd-programme} has funded research into how best
to support detailed practice in each of these areas (including this
present report), and that of the
non-science UK research councils.

Below, the RCUK principles are referred to as R$n$, and the STFC
principles and recommendations as SP$n$ and SR$n$ (these are
additionally reproduced in Appendix~\ref{s:stfc-principles}, for
convenience).

The STFC policy comprises
a number of \q{general principles} followed by some \q{recommendations for
good practice}.  There is no direct
linkage between the STFC policies and the RCUK principles, despite the
declaration as SP1 that \q{STFC policy incorporates
the joint RCUK principles on data management and sharing.} The
relationships given here are an interpretation by the 
authors of this report.  We also bring out some implications for data management plans
based on these policies.  Further implications were discussed by the
GridPP project~\cite{britton12} and the STFC Computing Advisory
Panel~\cite{stfc-cap12}.


\newenvironment{quotedprinciple}{\itshape \textbf{Principle:} }{\par}

\subsubsection{R1: data is a public good and should
  be shared}
\begin{quotedprinciple}
Publicly funded research data are a public good, produced in the
public interest, which should be made openly available with as few
restrictions as possible in a timely and responsible manner that does
not harm intellectual property.
\end{quotedprinciple}

Relates to STFC principles SP3, SP10, SP11 and SP12.
SP3 essentially defines what is meant by data,
distinguishing between \q{raw}, \q{derived} and \q{published}
data.  SP10 and SP11 acknowledge the need for an embargo
period\footnote{\q{Embargo period} is a better term than
  \q{\gls{proprietary period}}, though the latter term is conventional in
  some areas.}, while emphasizing the goal of public availability,
while principle S12 also introduces the possibility of registration to
track usage of data. Thus the STFC policy clarifies what restrictions
may be required and attempts to define more closely how they could be
implemented.  The stipulation that data should be shared is qualified
in~R4 and~R5 by a discussion of societal constraints, and professional embargo periods.

See further considerations for data management planning in
\prettyref{s:sharing} (on sharing) and
\prettyref{s:data-release-planning} (on planning).

\subsubsection{R2: projects should follow community best practice}
\begin{quotedprinciple}
Institutional and project specific data management policies and plans
should be in accordance with relevant standards and community best
practice. Data with acknowledged long-term value should be preserved
and remain accessible and usable for future research. 
\end{quotedprinciple}

Relates to SP5, SP6, SP7, SP8 and SP9, and recommendations SR1, SR4, SR5 and SR6.
The RCUK principle introduces the idea of a plan for data management,
one of whose aims is long-term access and usability of the data. The
STFC policy has much more to say about plans. Principle S5 requires
that they exist for data within scope, and principle S6 makes them
mandatory for grant-funded projects. Principles S7 and S8 also
make them required of STFC \glspl{facility} and desirable of external
facilities. Principle S9 echoes the RCUK emphasis on standards and
best practice.

The STFC recommendations offer advice on the relationship between
plans and facility policies, what data should be covered, and the
needs of long-term preservation. The Digital Curation Centre's
guidance is specifically mentioned. SR6\dash which should
perhaps be seen as a policy statement\dash asserts that original data
should be retained for ten years after the end of the project, and
non-reproducible data should be kept in perpetuity.  This has resource
implications, and so relates to~R7.

This principle more-or-less directly entails this present document, or
something like it, but more specific
implications for data management planning include the following
\begin{itemize}
\item We must distinguish data management planning for the \gls{facility}
  from data management planning for grants/projects that use the
  facility; this has an effect on the budgetary structure of the facility.
\item We should involve stakeholders in setting data retention and
  access policy.
\item Data management planning is part of the science funding
  lifecycle (and thus another link to~R7, and to \prettyref{s:preservation-costs}).
\item Best practice will be specific to each scientific domain.
\item The principle has implications on long-term preservation planning (ten years or more; see
  \prettyref{s:storagecosts} on the costs of long-term storage, and
  \prettyref{s:storage} for some dissection of the threats).
\end{itemize}

\subsubsection{R3: metadata should be available}
\begin{quotedprinciple}
To enable research data to be discoverable and effectively re-used by
others, sufficient metadata should be recorded and made openly
available to enable other researchers to understand the research and
re-use potential of the data. Published results should always include
information on how to access the supporting data. 
\end{quotedprinciple}

Relates to SR6, which recognizes that sufficient metadata is required to
enable reuse of data: to some extent this is addressed by the presence
of \glspl{Retrieval Aid} in the OAIS implementation, and to some
extent by the complexities of developing suitable \gls{Representation
  Information} as discussed elsewhere in this document (for example in
\prettyref{s:case-data-preservation} on sharing, and
\prettyref{s:oais-audit} on auditing).

The metadata within the repository need not be the only metadata
available, nor even necessarily the best.
Some biological data repositories, notably
Dryad,\footnote{\url{http://www.datadryad.org/}} set only minimal
(DataCite-compliant) metadata requirements for data
deposits,\footnote{The DataCite project \url{http://datacite.org/} is prominent in the effort to associate DOIs with datasets}
on the grounds that the
deposited datasets are associated with a peer-reviewed journal
article, and that it is this article which provides the best
human-readable information.  While this usefully avoids extra effort for the
data producer, it is in tension with the OAIS principle that the archive
should take responsibility for (which here means control over) all
aspects of the \gls{AIP}.  The resolution has to be a pragmatic one,
and may involve extraction of metadata from, or wholesale inclusion
of, the associated article, which of course brings in both
technological and copyright problems.

Implications for data management planning include:
\begin{itemize}
\item sufficiency and availability of metadata;
\item relationship to \gls{OAIS} (\gls{Representation Information} etc.);
\item how to link from publications to data (the question of data
  citation is a large one, which we do no more than touch on in this report).
\end{itemize}

\subsubsection{R4: legitimate constraints on release}
\begin{quotedprinciple}
RCUK recognises that there are legal, ethical and commercial
constraints on release of research data. To ensure that the research
process is not damaged by inappropriate release of data, research
organisation policies and practices should ensure that these are
considered at all stages in the research process. 
\end{quotedprinciple}

Relates to  SP2.
R4 is the other side of the coin to~R1, where
public good had primacy.  SP2 is a terse
acknowledgement of the need to comply with relevant legislation.

Implications for data management planning include
commercial confidentiality,
data protection, and 
freedom of information.  See \prettyref{s:data-release-planning}.

\subsubsection{R5: researchers are entitled to some privileged use}
\begin{quotedprinciple}
To ensure that research teams get appropriate recognition for the
effort involved in collecting and analysing data, those who undertake
Research Council funded work may be entitled to a limited period of
privileged use of the data they have collected to enable them to
publish the results of their research. The length of this period
varies by research discipline and, where appropriate, is discussed
further in the published policies of individual Research Councils. 
\end{quotedprinciple}

Relates to SP10 and SP11.
R5 is a further qualifier on~R1, this time
from the perspective of academic reward to those who have collected
the data. The STFC principles expresses this in similar terms, but
with an expectation that \q{published} data should generally be
available within six months of the date of the publication.

Implications for data management planning include defining and
implementing embargo periods, and this again comes under the catch-all
remit of \prettyref{s:data-release-planning}.

\subsubsection{R6: data use should be acknowledged}
\begin{quotedprinciple}
In order to recognise the intellectual contributions of researchers
who generate, preserve and share key research datasets, all users of
research data should acknowledge the sources of their data and abide
by the terms and conditions under which they are accessed. 
\end{quotedprinciple}

This is not explicitly referred to in the STFC policy, perhaps on the
grounds that it appears to be a statement of normal academic good
practice.  However it is a nod towards the importance of the ongoing
work on developing the technical infrastructure for data
citation (for example DOIs for datasets).



\subsubsection{R7: DMP planning should be funded}
\begin{quotedprinciple}
It is appropriate to use public funds to support the management and
sharing of publicly-funded research data. To maximise the research
benefit which can be gained from limited budgets, the mechanisms for
these activities should be both efficient and cost-effective in the
use of public funds. 
\end{quotedprinciple}

The obligation here is on the funders to support the activities which
their principles demand, but the extent and cost of support must be
negotiated with funded projects.  Since the data's \glspl{Designated
  Community} will include both professionals and the wider society,
the discussion of what is a minimally acceptable preservation strategy
must be negotiated as well.

This is obliquely referred to in SR6, where
\q{[i]t is recognised that a balance may be
  required between the cost of data curation (eg for very large
  data sets) and the potential long term value of that data.}  See
also the discussion of costs in \prettyref{s:preservation-costs}

\subsubsection{Other STFC principles}
A number of STFC principles and recommendations do not appear to
derive from or relate directly to the RCUK principles. These are SP4,
on STFC's reponsibilities for data use, SP13 on data integrity, SR2 on
choice of repositories and SR3 on quality assurance of data products.

Implications for data management planning include:
the choice of repository (where this is not obvious),
the development and maintenance of a provenance trail, and
integrity checking.

\begin{takeaways}{planners}
\item \gls{STFC} has articulated a set of high-level principles
  governing scientific data management, and is currently (August 2012)
  consulting on these principles with relevant stakeholders in the HEP
  community.
\end{takeaways}

\subsection{Sharing: openness and citation}
\label{s:sharing}

\subsubsection{The argument for open data}
\label{s:openness}
\index{open data|(}
Internationally, there is a push towards such
\gls{data sharing} in the more general context of scholarly research
(see for example~\cite{Arzberger2004} or~\cite{ruusalepp08}).  

We have already discussed the \gls{STFC} data sharing principles.
Regarding publications, STFC, in common with the other UK research councils, requires
that
\begin{quotation}
the full text of any articles resulting from the grant that
  are published in journals or conference proceedings [\dots] must be
  deposited, at the earliest opportunity, in an appropriate e-print
  repository\cite[\S8.2]{stfc-rgh}.\footnote{\url{http://www.scitech.ac.uk/rgh/rghDisplay2.aspx?m=s&s=64}}
\end{quotation}
The \gls{RCUK} policy goes further, and mandates (from April 2013) that research
funded by the (UK) Research Councils \qq{must be published in journals
  which are compliant with Research Council policy on Open
  Access}~\cite{rcuk-outputs}, which requires publication through either Gold Open Access
(an open-access journal) or Green Open Access (the journal permits
self-archiving).\footnote{\url{http://en.wikipedia.org/w/index.php?title=Open_access&oldid=502798213}}

In the US, the \gls{NSF}'s GC-1
document~\cite{nsf-gc1} states in section~41 that \qq{[NSF]
  expects investigators to share with other researchers,  at no more
  than incremental cost and within a reasonable time, the data,
  samples, physical collections and other supporting materials created
  or gathered in the course of the work. It also encourages grantees
  to share software and inventions or otherwise act to make the
  innovations they embody widely useful and usable.} This is
reiterated in almost the same words in their 2010 data sharing policy~\cite{nsf11}.
They additionally require a brief statement, attached to proposals, of
how the proposal would conform to NSF's data-sharing policy.

The year 2009 saw some excitement (arising from the incident inevitably
labelled \q{climategate}, and to some other data-release
disputes\footnote{\url{http://www.guardian.co.uk/environment/2010/apr/20/climate-sceptic-wins-data-victory}})
related to the management and release of climate data.\index{climate
  data} This illustrated the political and social significance of some
science data sets; the contrast between what scientists know, and the
public believes, to be normal scientific practice; and some of the
issues involved in the generation, ownership, use and publication of
data.\footnote{UEA's Climate Research Unit is a partner in the ACRID
  project, also funded by the JISC \gls{MRD}
  programme: \url{http://www.cru.uea.ac.uk/cru/projects/acrid/}}  The cases during that year
illustrate a number of complications involved in data releases.
\begin{enumerate}
\item Data is often passed from researchers or groups directly to
others, across borders, with no general permission to distribute it
further.
\item Data collection may be onerous, and the result of significant
professional and personal investments.
\item Raw data\index{raw data} is generally useless without the more
or less significant processing which cleans it of artefacts and makes
it useful for further analysis.
\item However not all disciplines have the clear notion of published \gls{data
products}\index{data products} which is found in astronomy and which is
implicit in the \gls{OAIS} notion of archival deposit.\label{item:dataproducts}
\item Science is a complicated social process.
\end{enumerate}
In science, we preserve data so that we can make it available
later. This is on the grounds that scientific data should generally be
universally available, partly because it is usually publicly paid for,
but also because the public display of corroborating evidence has been
part of science ever since the modern notion of science began to
emerge in the 17th century\dash witness the Royal Society's
motto, \q{nullius in verba}, which the Society glosses as \q{take
  nobody's word for it}. Of course, the practice is not quite as
simple as the principle, and a host of issues, ranging across the
technical, political, social and personal, complicate the social,
evidential and moral arguments for general data release.

The arguments \emph{against} general data releases are
practical ones: data releases are not free, and may have significant
financial and effort costs (cf \prettyref{s:preservation-costs}).\index{costs}
Many of these costs come from
(preparation for) data preservation, since it is formally archived data
products that are the most naturally releasable objects: releasing raw or low-level
data \emph{may} be cheap, but may also have little value, since raw
underdocumented datasets are likely to be
useless; or more pessimistically such data releases may even have a negative value, if they
end up fostering misunderstandings which are time-consuming to counter
(this point obviously has particular relevance to politicised areas such as
climate science).
In consequence of this, the \q{open data question} overlaps with the
question of data preservation\dash if the various costs and sensitivities
of data preservation are satisfactorily handled, then a significant
subset of the practical problems with open data release will promptly
disappear.  We discuss the data preservation question below, in
\prettyref{s:case-data-preservation}.

Some questions of data sharing can be usefully discussed using the
OAIS notion of the \gls{Designated Community} and the associated
\gls{Representation Information} that the Community is expected to
find intelligible. Higher level \gls{data products} contain less
detail than lower-level or raw datasets; they are also intended to
serve broader Communities, and are more expensive to generate in terms
of processing and QA.  We have no data about the costs of
documentation, but we suspect that rawer data is more expensive to
document than higher-level data products.  When a scientist chooses
between a project's available data
products, the choice will represent a trade-off involving the amount
of time they can afford to invest in understanding the data product
(via its Representation Information), the degree of support they can
hope to receive from colleagues and the data owners, and the subtlety
of the question they wish to answer (more subtle distinctions might be
erased by higher-level products, but might be spuriously detected in
poorly-understood rawer data).  On the other side of the exchange, a
project will have a formal or informal model of whom it is serving by
the provision of data, and will design data products, and allocate
costs, accordingly.

It seems worth noting, in passing, that the physical sciences broadly
perform better here than other disciplines, both in the technical
maturity of the existing archives and in the community's willingness
to allocate the time and money to see this done effectively.

\index{open data|)}

\subsubsection{The argument for data preservation}
\label{s:case-data-preservation}

As an observational science, astronomy data is generally repeatable,
but some of the most precious astronomical data records unpredictable
transient events or (through historical observations) long-timescale
secular changes.  Astronomical data is potentially useful almost
indefinitely and, because its object of study is in some sense
fundamentally simple (there is only one sky, after all), it is also
broadly intelligible almost indefinitely.

\gls{HEP} data is somewhat different.
As an experimental science, it is generally very much in control of
what it observes through the successive generations of experiments it designs.
A consequence of this is firstly that HEP experiments have a much
stronger tendency to become obsolete with each technological
generation, and secondly that the complication of the apparatus makes
it hard to communicate into the future a level of understanding
sufficient to make plausible use of the data.  Experimental apparatus
will generally be understood better and better as time goes on (this
is also true of satellite-borne detectors in astronomy), so that data
gathered early in an experiment will be periodically reanalysed with
increased accuracy.  However this understanding is generally not
preserved formally, but is pragmatically communicated through wikis,
workshops, word of mouth, configuration and calibration files, and
internal and external reports.  Even if all of the tangible records
were magically preserved with complete fidelity, and supposing that
the more formal records do contain all the information required to
analyse the raw data, an archive would still be missing the
word-of-mouth information which a new postgrad student (for example)
has to acquire before they can understand the more complete
documentation.  We can think of this as a \q{bootstrap problem}.  In
OAIS terms, the \gls{Representation Network}
for \gls{HEP} data is particularly intricate, and while the
\gls{Representation Information} nearest to the \gls{Data Object} may
be complete, it may be infeasible to gather the Representation
Information necessary to let a naive researcher make sense of it.  The
\gls{Designated Community} for \gls{HEP} data may therefore be null in the
long term.

This sounds pessimistic, but~\cite{south11} describes a number of
scenarios in which \gls{HEP} data can and should be reanalysed some decades
after an experiment has finished, and describes ongoing work on the
development of consensus models for preserving data for long enough to
enable such post-experiment exploitation.  This provides a case
for a style of preservation somewhat different from the astronomical
one.  What these scenarios have in common is a commitment of a few FTEs
of staff to actively conserve and continuously exploit the data.  This
post-experiment staff can therefore 
be conceived as a form of walking \gls{Representation Information} so that,
while they are still involved, the data might have a \gls{Designated
Community} which corresponds to those individuals in a position to
undertake an extended apprenticeship in the data
analysis.

Finally, and as noted in \prettyref{s:openness}, if data is well
archived, then most of the pragmatic objections to opening that data
do not apply (though not the professional-credit reasons).  Thus, to
the extent that general data release is a good 
in itself, it is a further argument in favour of a well supported archive.

\subsubsection{Should everything be preserved?}
\label{s:preserve-raw}

In the data-preservation world, there is often an automatic expectation that \q{everything
  should be preserved}, so that an experiment can be redone, results reanalysed, or an
analysis repeated, later.  Is this actually true?  Or if it is at
least desirable, how much effort should be expended to make it true?
This question is implicit in, for example, the discussion of software
preservation\index{software preservation} in
\prettyref{s:sw-preservation}.

In fact, it is not always the case that an experiment can feasibly be
redone, because it is not always feasible to document an experiment in
enough detail that the measurements can be remade.   For similar
reasons, if the data analysis is particularly complicated, or requires
a particularly subtle understanding of the behaviour of a particular
instrument, it may not be feasible to document that analysis in enough
detail that the data can be reanalysed.
There is therefore a case that at least some details of the
experimental environment\dash digital as well as physical\dash are not
reasonably preservable, and that as a result
little effort should be expended on preserving them, if
well-documented higher-level data products are available and intelligible.


We should stress that we are not advocating deliberately deleting raw
data, and its associated pipelines\dash it \emph{might} be useful, and it
\emph{might} be usable\dash but simply noting that one should not
overstate its value.

This argument is examined in a little more detail in~\cite[\S2.4]{gray10}.

\begin{takeaways}{planners and funders}
\item Sharing data is generally agreed to be a virtue, but it should
  not be regarded as trivial, and it may incur significant costs and complications.
\item Many of the problems are directly or indirectly practical
  problems, to do with required resources; but well described data, ready
  for long-term preservation, is as a side-effect easily shared data,
  so that solving the preservation problem can also partially address
  the pragmatics of data sharing.
\item Data must be documented fully enough that it can be used by
  its intended audiences, whoever that is.  If it is not so documented, it is probably
  not worth releasing, and indeed releasing it may be harmful overall.
\end{takeaways}

\section{Technical background}
\label{s:technical}

This section is concerned with the various technical frameworks
relevant to the good management of data.  None of these frameworks is
of a type which can be mechanically applied to a given preservation
problem -- there are no turnkey solutions here -- but we include
these topics to illustrate the range of technical developments, as
opposed to policy issues of \prettyref{s:policy} and the practical
planning actions of \prettyref{s:planning}, which might be of interest to the developer of
a preservation plan.


\subsection{OAIS}
\subsubsection{Description}
\label{s:oais-description}
The discussion in this document is structured around the OAIS model. We
introduce here the main concepts of the OAIS model. Full details are
in~\cite{std:oais} with useful introductory guides in~\cite{lavoie04}
and \cite[chs.3 \& 6]{giaretta11},
and some discussion in the LSC context in~\cite{anderson11}. 

The term \emph{OAIS} stands for an \emph{Open Archival Information
  System}.  The word \q{open} is not intended to imply that the archived
data is freely available (though it may be), but instead that the
process of defining and developing the system is an open one.
The principal concern of an OAIS is to
preserve the usability of digital artefacts for a pragmatically
defined long term.  An OAIS is not only
concerned with storing the lowest-level \emph{bits} of a digital
object (though this part of its concern, and is not a trivial problem), but with storing enough
\emph{information} about the object, and defining an adequately
specified and documented \emph{process} for migrating those bits from
system to system over time, that the information or knowledge
those bits represent can be retrieved from them at some indeterminate
future time.  The OAIS model can therefore be seen as addressing an
administrative and managerial problem, rather than an exclusively
technical one.

\begin{figure}
\begin{centering}
\includegraphics[width=0.75\columnwidth]{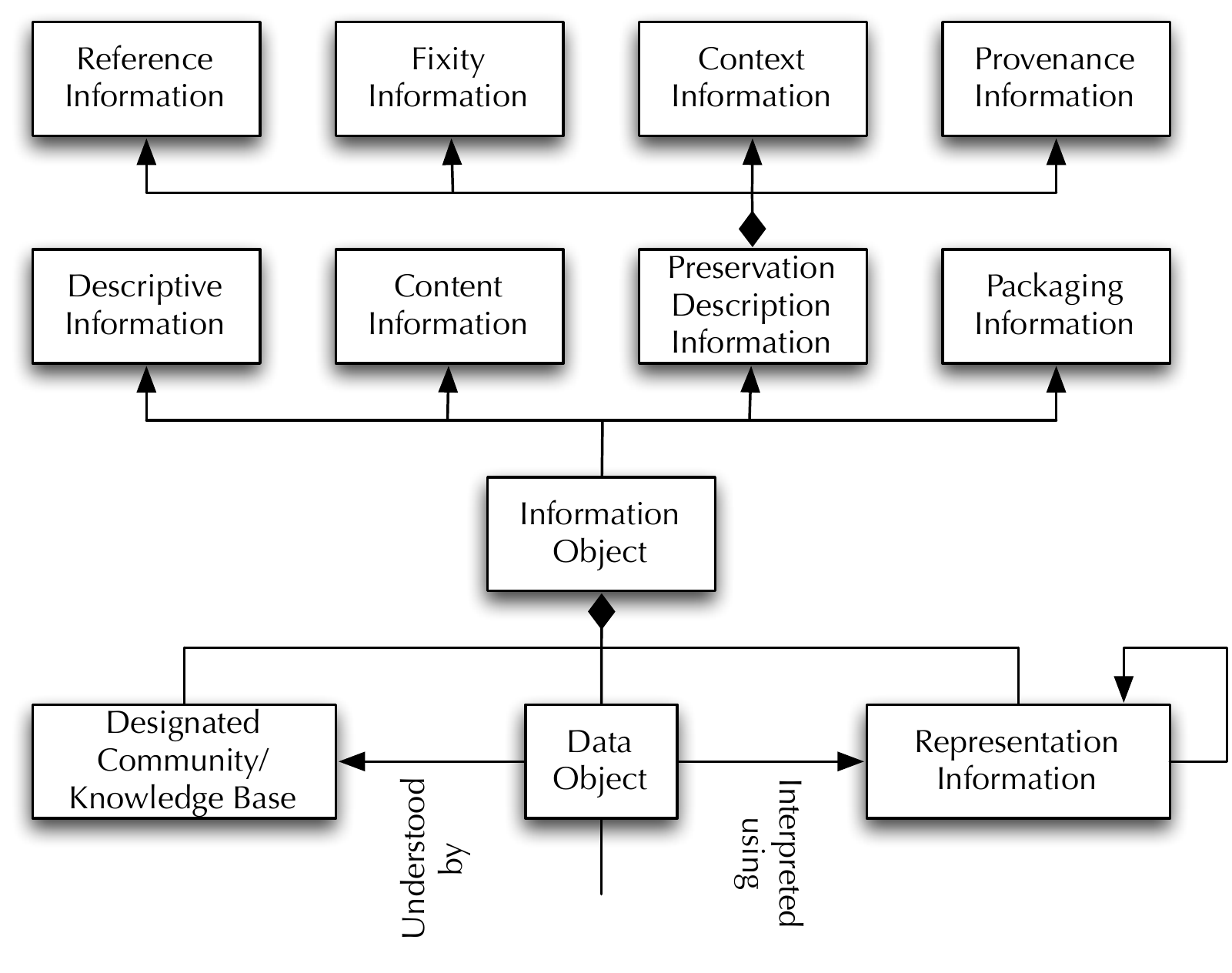}\\
\end{centering}
\caption{\label{f:oais-brian}The OAIS information model.  The \q{data
    object} is the bag of bits which is being preserved}
\end{figure}

\begin{figure}
\includegraphics[width=\textwidth]{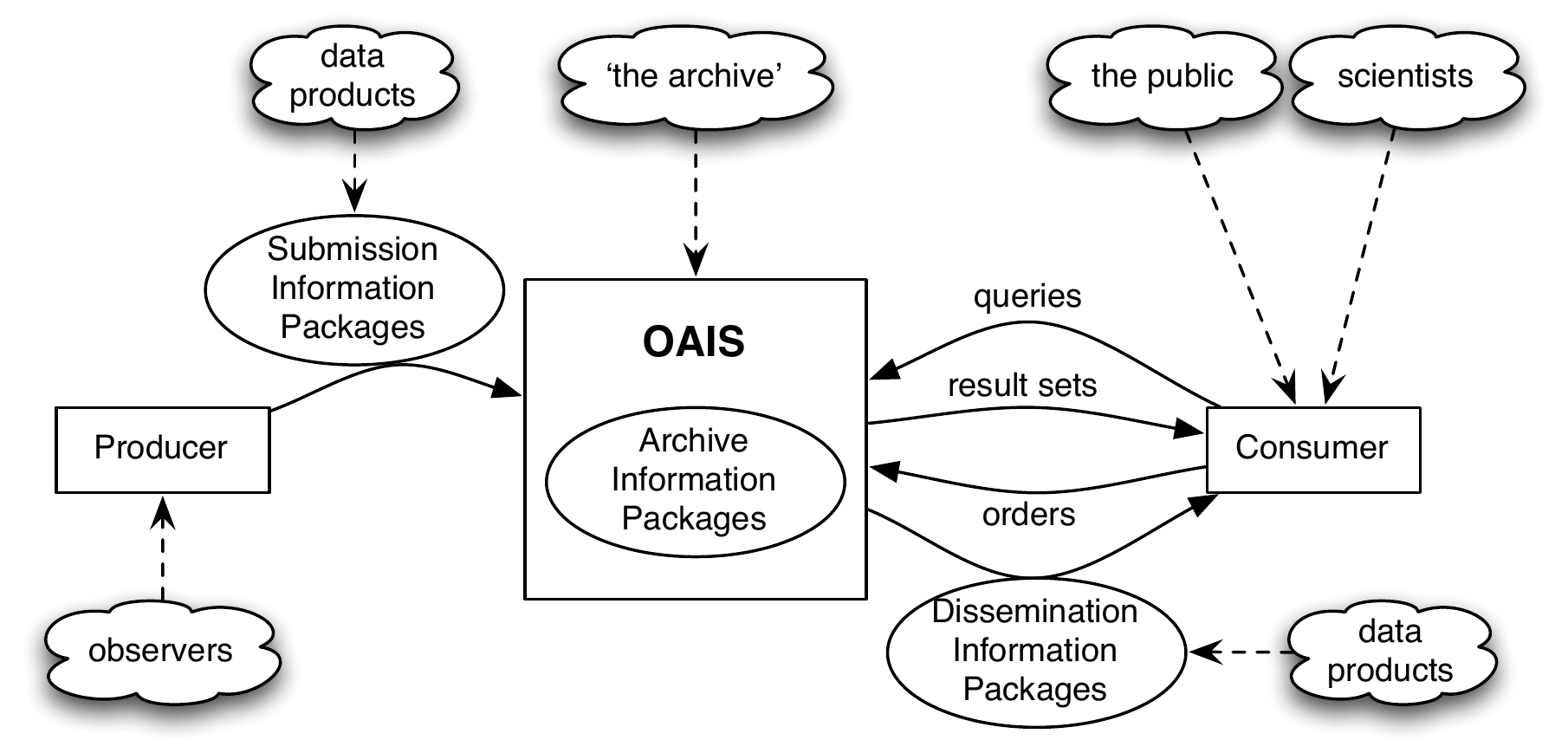}
\caption{\label{f:oais-annotated}The highest-level structure of an
  OAIS archive, annotated with the corresponding labels from
  conventional astronomical practice (redrawn from~\cite[Fig.~2-4]{std:oais}).
  The dissemination data products will often in practice be the same as the
  submitted ones, but archives can sometimes create value-added ones
  of their own.}
\end{figure}

The OAIS specification's principal output is the \emph{OAIS reference
  model}, which is an explicit (but still rather abstract) set of
concepts and interdependencies which is believed to exhibit the
properties that the standard asserts are important.  The structure of
the information model is illustrated in \prettyref{f:oais-brian}, and
the structure of the relationships between Producers and Consumers in
\prettyref{f:oais-annotated}.

\begin{figure}
\begin{centering}
\includegraphics[width=0.75\columnwidth]{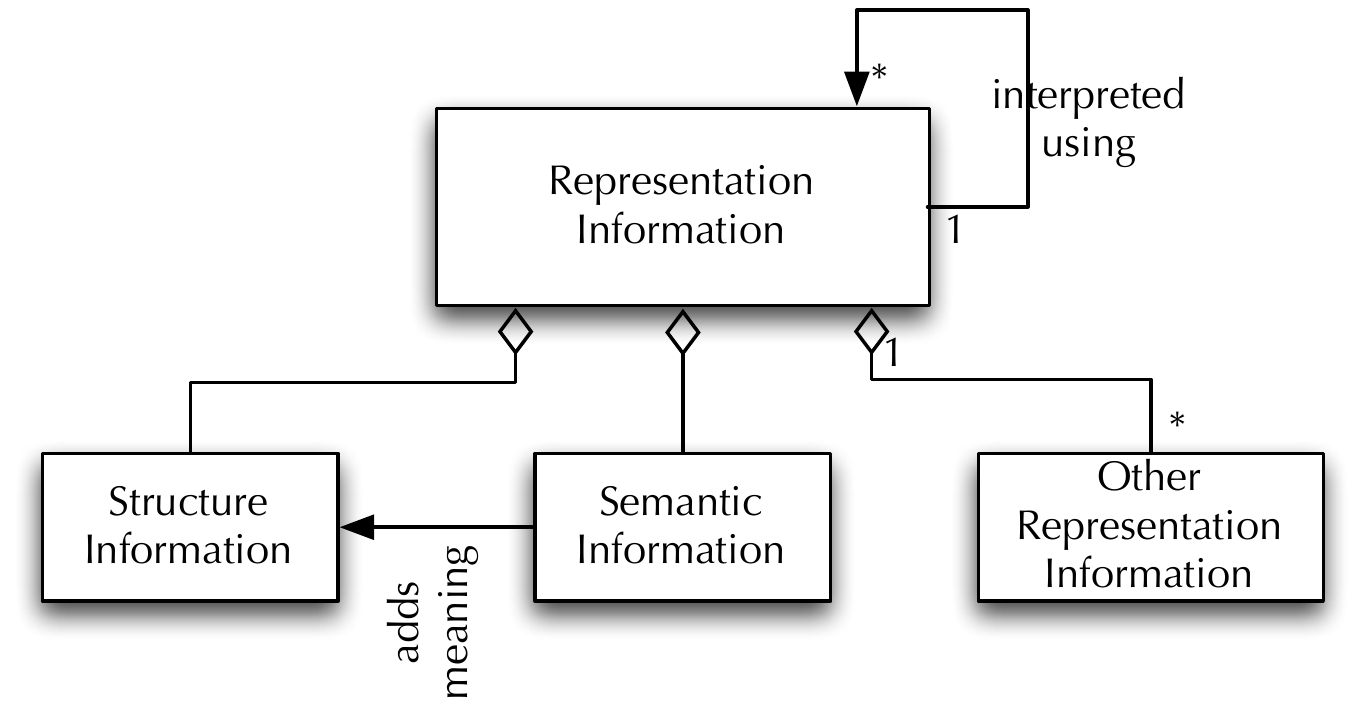}\\
\end{centering}
\caption{\label{f:oais-rio}Representation Information Object}
\end{figure}

An OAIS archive is conceived as an entity which preserves objects
(digital or physical) in the \gls{Long Term}, where the \q{Long Term} is
defined as being long enough to be subject to technological change.  The
archive accepts objects along with enough \gls{Representation
  Information} to describe how the digital information in the object
should be interpreted so as to extract the information within it (for
example, the FITS specification is Representation Information for a
FITS file, or the NeXus specification for a NeXus file, in either case
accompanied by a dictionary which defines the meaning of keywords not
included in the underlying standard).   That Information may need
further context\dash for example, to document the PDF format of a
specification, or even to document what \q{ASCII}
means\dash and the collection of such explanations turns into a
\gls{Representation Network}, as illustrated in \prettyref{f:oais-rio}.
This information is all submitted to the archive in the form
of a \gls{SIP} agreed in some more or less
formal contract between the archive and its data \glspl{Producer}.

Once the information is in the archive, the long-term responsibility
for its preservation \emph{is transferred from the Producer to the archive},
which must therefore have an explicit plan for how it intends to
discharge this.  No matter how closely related are the archive and the
data Producer, the transfer reflects the extent to which the archive has
different goals and timescales from the day-to-day management of the
working data.

The archive preserves its contents in the form of \glspl{AIP}, and
distributes them to \glspl{Consumer} in one or more
\glspl{Designated Community}
by transforming them into the \gls{DIP} which corresponds to
a \q{data product}.\footnote{In practice, there may be only minor
  differences between the data products forming SIPs, AIPs and DIPs,
  and the differences will generally have more to do with management
  metadata than physical content.}  The members of the Designated Community are those
users, in the future, whom the archive is designed to support.  This
design requires including, in the \gls{AIP},
Representation Information at a level which allows the Designated
Community to interpret the data products \emph{without ever having met
  one of the data \glspl{Producer}}, who are assumed to have died, retired,
or forgotten their email addresses.

\begin{takeaways}{planners}
\item The OAIS vocabulary is a coherent, principled and shared vocabulary for archive planning.
\item OAIS is not concrete enough to support detailed planning by
  itself.
\end{takeaways}
\begin{takeaways}{funders}
\item The conversation with projects can be conducted in OAIS terms.
\item OAIS provides a framework for negotiating the archiving aspects
  of project costs/support.
\end{takeaways}

\glsadd{Information Package}
\index{OAIS|)}

\subsection{Preservation Analysis in CASPAR}
\label{s:caspar}

As we have noted above, the OAIS model is useful but somewhat vague.
The CASPAR project is an attempt to concretise the model with both a
more detailed analysis methodology, and a set of software tools.
CASPAR was a large-scale project in digital preservation funded under
the European Commission's 6th Framework Programme, bringing together
17 partners working on research, standards, policy development and
applications, and led by STFC.  For a summary
see~\cite{caspar-handbook}.

\begin{wrapfigure}{O}{4cm}
\includegraphics[width=4cm]{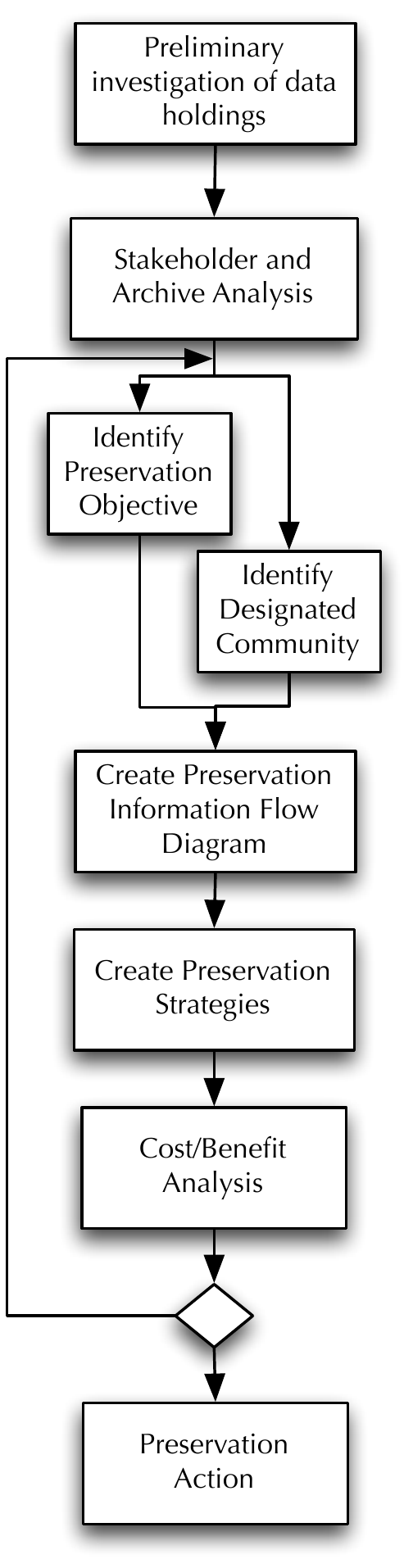}
\caption{\label{f:preservation}Preservation analysis workflow, from~\cite{conway11}}
\end{wrapfigure}

The aim of CASPAR was to develop the notion of an archival
information system as specified by OAIS and develop a set of methods
and tools for several stages of the digital preservation
lifecycle. There were three test beds in the project, in the domains
of cultural heritage, performing arts and science data, providing
demanding validation of the developments within the project.  The
science data test bed was provided by STFC and the European Space
Agency.  The output of the project is collected together
in~\cite{giaretta11}.  We consider two aspects of this project:
preservation analysis, and the preservation toolkit.

Validation is an alternative way of approaching the problem, which we discuss in
\prettyref{s:oais-audit}.

\subsubsection{A  preservation analysis approach}
\label{s:preservation-analysis}

As part of the work of CASPAR and some related case studies, a
preservation analysis method was developed~\cite{conway11,conway11a}.
This method is designed to ensure that the science data stored in the
archive is a truly reusable asset, capitalizing on a community's
expertise and knowledge by appreciating the nature of data use,
evolution and organizational environment.  It seeks to design the
optimal asset by capturing key information which allows reuse.  A
judicious analysis permits the design of \glspl{AIP} which deliver a
greater return of investment by both improving the probability of the
data being reused and potential outcome of that reuse.

The methodology incorporates a number of analysis stages into an
overall process capable of producing an actionable preservation plan
for scientific data, which satisfies a well defined preservation
objective. The challenge of digitally preserving scientific data lies
in the need to preserve not only the dataset itself, but also the
ability it has to deliver knowledge to a future user community. This
entails allowing future users to re-analyze the data within new
contexts. Thus, in order to carry out meaningful preservation, we need
to ensure that future users are equipped with the necessary
information to re-use the data.

The methodology specifies a number of stages in an overall process to
produce an actionable preservation plan for scientific data
archives. \prettyref{f:preservation} illustrates the process.  We
briefly discuss these stages here.  Although these analyses may at
first seem burdensome, we expect that since large-scale science
projects will have, or will need to develop, highly functional data
management systems; this means many of the questions below will
already have answers available in the data management design
documents, and other technical personnel already involved in the project.

\textbf{1. Preliminary Investigation of Data Holdings.}  A preliminary
investigation of the data holdings of the archive to: understand the
information extracted by users from data; identify likely Preservation
Description and \gls{Representation Information}; and develop a clearer
understanding of the data and what is necessary for its effective
re-use.  The CASPAR project developed a questionnaire which allowed
the preservation analyst to initiate discussion with the archive.

\textbf{2. Stakeholder and Archive Analysis.}  A stakeholder analysis
to identify: the producers of the data; the custodians of the data;
the custodians of other information required for reuse; the end users
groups.   Each stakeholder may hold different views of the knowledge a
data set provides. It is also beneficial to understand how an archive
has evolved and been managed to uncover different uses of data over
time.

\textbf{3. Defining a Preservation Objective.}  One or more
preservation objectives should be identified which are: well defined
and clear to anyone with a basic knowledge of the domain; currently
achievable; and can be assessed to determine when the objective has
been attained by the adopted preservation strategy.

\textbf{4. Defining a Designated User Community.}
An archive defines the Designated
Community for which it is guaranteeing to preserve some digitally
encoded information, and that Community possesses the skills and
knowledge to use the information within an \gls{AIP} in order to
understand and reuse the data. In common with the preservation
objective, there may be a range of community groups that the archive
may chose to serve. The definition of the skills is vital, as it
limits the amount of information which needs to be contained within an
AIP in order to satisfy a preservation objective.  

\textbf{5. Preservation Information Flows and Strategies.} Once the
objective and community have been identified, the information required
to achieve an objective for this community can be determined, and planners
can develop the appropriate AIPs. OAIS specifies that within
an archival system, a data item has a number of information items
associated with it. The preservation objective should be satisfied
when each item of the OAIS information model has been adequately
populated. The information model thus provides a checklist which
ensures that the preservation objective can be met, and determines the
strategies available to meet that objective, as alternative
information items may be available to meet the objective.   Multiple
strategies can thus be developed, each specifying a series of clear
preservation actions in order to create an AIP. 

\textbf{6. Cost/Benefit/Risk Analysis.}  The final stage of the
workflow is where plan options can then be assessed according to:
costs to the archive directly, as well as the resources knowledge and
time of archive staff; benefits to future users which ease and
facilitate re-use of data; risks inherent to the preservation
strategies and accepted impact to the archive.

Once this analysis is complete, the optimal strategy can be selected
and progressed to preservation action within the archive.  

Identifying the preservation information flows and strategies is
perhaps the most technically involved step of this process.  As a
consequence, CASPAR and subsequent projects have developed the notion
of a \gls{PNM} as a tool to analyse the preservation information and
strategies available to the archive.    A PNM is a formal
representation of the digital objects under consideration, which
allows a preservation objective to be met for a future designated
community.  It identifies the dependencies between a digital object
and its related Representation Information, and includes the
alternative approaches to satisfying the preservation objective.   A
network can then be traversed to estimate the costs and risks
associated with a particular strategy.   Work on using PNM is ongoing
in the European projects SCAPE and
SCIDIP-ES\footnote{SCAlable Preservation
  Environments \url{http://www.scape-project.eu/}  and SCIence Data Infrastructure for Preservation\dash
  Earth Science \url{http://www.scidip-es.eu/}}, including some initial
analysis of digital assets of the ISIS \gls{facility}~\cite{conway12}.

\begin{takeaways}{planners}
\item There exists a semi-standard procedure for developing a DMP
  plan.
\item Pre-existing data management design documents should make this process more lightweight
  than it may at first appear.
\end{takeaways}

\subsubsection{The CASPAR Toolkit}

The preservation toolkit developed an integrated architecture and
tools to support the various phases of the preservation process as
described in OAIS functional model.  These include:
\begin{itemize}
\item Representation Information Toolkit: to aid the identification,
  creation, maintenance and reuse of OAIS Representation Information.
\item Registry of representation information: Centralised and
  persistent storage and retrieval of OAIS Representation Information,
  including Preservation Description Information.
\item Packaging tools: the construction and un-packaging of OAIS
  Information Packages.
\item An approach to the authenticity of digital objects: the
  maintenance and verification of authenticity in terms of identity
  and integrity of the digital objects.
\item Virtualisation services:  to allow the search for an object using either a related measurable parameter or a linkage to remote values.
Knowledge management for preservation planning: these allowed the
definition of Designated Communities, and the identification of
missing Representation Information.
\item Orchestration Services: the reception of notifications of
  changes events which impact preservation, triggering preservation
  actions to respond to these changes and sending of alerts to
  Subscribers.
\item Access and rights management: the definition and enforcement of
  access control policies, and the registration of provenance
  information on digital works and retrieval of rights holding
  information.
\end{itemize}
These tools and their interactions were in at a prototype stage at the
end of CASPAR; their development is being continued in the SCIDIP-ES
project. 

\subsection{Audit and certification of trustworthy digital repositories}
\label{s:oais-audit}
There has long been a recognised need for reliable and comprehensive
assessment
of digital repositories, measuring the degree to which they can be
trusted to preserve their contents into the future and maintain access
and usability. It is natural that such an assessment should be founded
on the OAIS as the international standard that sets out fundamental
requirements for a repository for long-term preservation.  After the
OAIS standard was produced, work continued\dash led by RLG/\gls{OCLC}
and the \gls{NARA}\dash towards a standard for accreditation of
archives. This resulted in the \q{\gls{TRAC}} document~\cite{dale07} which
was subsequently developed by a \gls{CCSDS} working group through a
public process,%
\footnote{See \cite[ch.25]{giaretta11} and \longurl{http://www.digital~repository~auditand~certification.org/}}  
and
taken into ISO in the same way that OAIS itself was.

The standard \q{Audit and certification of trustworthy digital
  repositories}~\cite{std:ccsds652.0} (CCSDS 652.0 = ISO-16363:2012)
was published in February 2012. It offers a detailed specification of
criteria by which digital repositories can be audited. Its scope is
the entire range of digital repositories.

The standard is grounded in OAIS and is intended to be completely
comprehensive.  It presents a series of metrics under the following
main headings:
\begin{itemize}
\item Organizational Infrastructure
\item Digital Object Management
\item Infrastructure and Security Risk Management
\end{itemize}
Each metric is accompanied by discussion and examples of how a
repository can show it is meeting the requirement expressed in the
metric. A typical example is shown in \prettyref{f:ccsds652.352}.

\begin{figure}
\fbox{%
\advance\columnwidth -2\fboxsep
\advance\columnwidth -2\fboxrule
\begin{minipage}\columnwidth
\parskip\medskipamount

\smallskip
\leftskip=1em
\rightskip=1em plus 3em

\textbf{3.5.2 The repository shall track and manage intellectual property rights and
restrictions on use of repository content as required by deposit agreement, contract, or
license.}

\textbf{Supporting Text}:
This is necessary in order to allow the repository to track, act on, and verify rights and
restrictions related to the use of the digital objects within the repository.

\textbf{Examples of Ways the Repository Can Demonstrate It Is Meeting
  This Requirement}:
A Preservation Policy statement that defines and specifies the repository's requirements and
process for managing intellectual property rights; depositor agreements; samples of
agreements and other documents that specify and address intellectual property rights;
documentation of monitoring by repository over time of changes in status and ownership of
intellectual property in digital content held by the repository; results from monitoring,
metadata that captures rights information.

\textbf{Discussion}:
The repository should have a mechanism for tracking licenses and contracts to which it is
obligated. Whatever the format of the tracking system, it must be sufficient for the institution
to track, act on, and verify rights and restrictions related to the use of the digital objects
within the repository.
\medskip
\end{minipage}}
\caption{\label{f:ccsds652.352}An example of repository metrics:
  section 3.5.2 of CCSDS 652.0~\cite{std:ccsds652.0}}
\end{figure}

It is expected that the standard will become widely used for auditing
digital repositories, and that services will be offered just as they
are for ISO 9000 and other standards-based certifications. There is an
associated standard under development \q{Requirements for bodies
  providing audit and certification of candidate trustworthy digital
  repositories}~\cite{std:ccsds652.1}. This allows for the
accreditation of organizations that will offer audit and certification
services.

We have more to say about the practicalities of validation in
\prettyref{s:practice-validation}.

\begin{takeaways}{planners}
\item There is ongoing work, plus some standardised conclusions in the
  form of CCSDS 652.0~\cite{std:ccsds652.1}, on how
  to extend OAIS to make it more concrete.
\end{takeaways}

\subsection{The DCC curation lifecycle model -- a contrast to OAIS}
\label{s:dcc-lifecycle}

The OAIS model is on the face of it a linear one, and suggests that
data is created, then ingested, then preserved, and then accessed, in
a process which has a clear beginning and end. This is compatible with
the observation that one point of archiving data is to reuse or
repurpose it, creating new archivable data products in turn, but this
longer-term cycle remains only implicit in the model. The OAIS model
is therefore very usefully explicit about those aspects of archival
work concerned with long-term preservation, but its conceptual
repertoire is such that a discussion framed by it runs the risk of
underemphasizing the range of roles a data repository has, or even of
marginalising it.

%
\begin{figure}
\begin{centering}
\includegraphics[width=8cm]{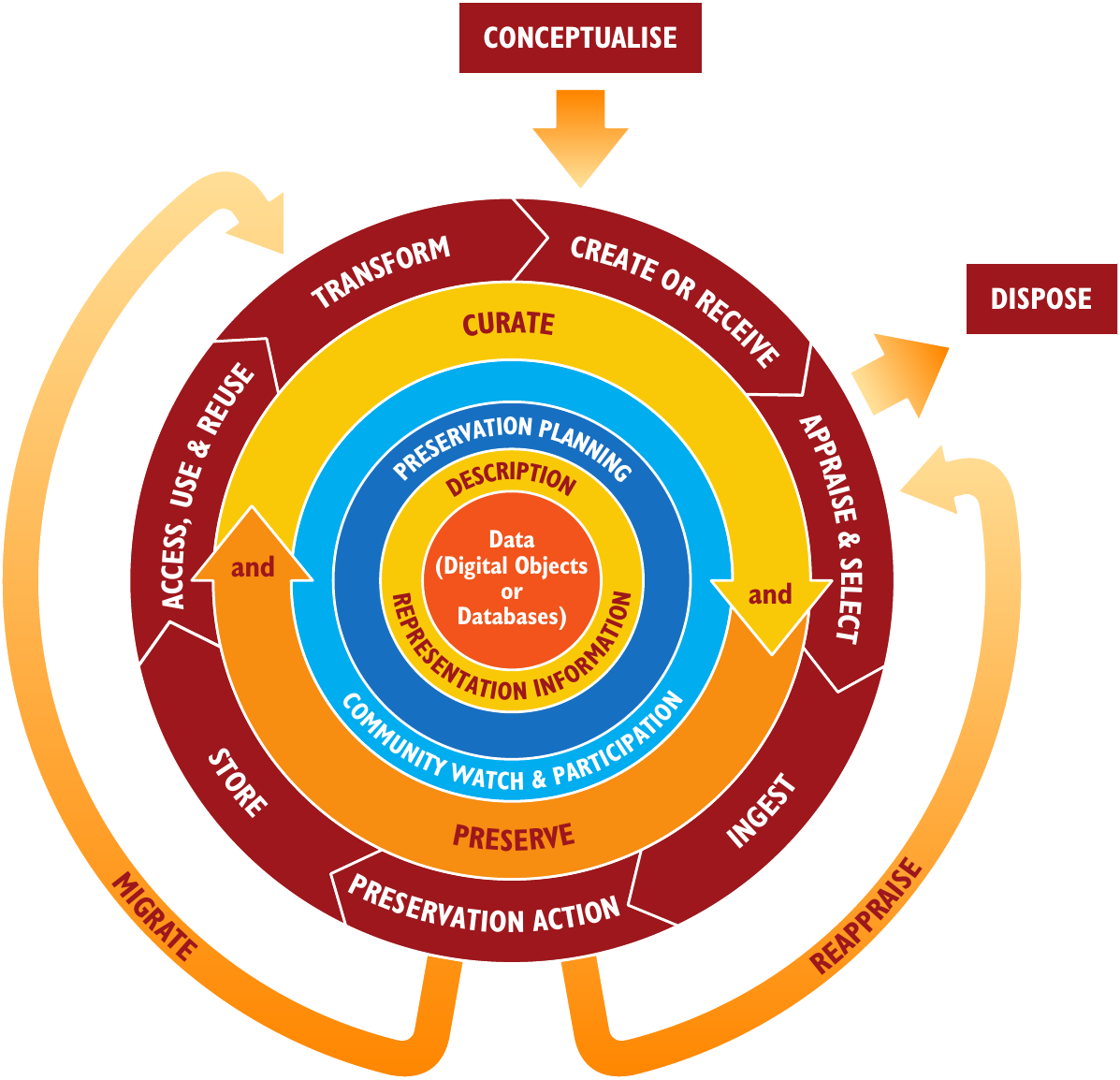}\\
\end{centering}
\caption{\label{f:dcc-lifecycle}The DCC lifecycle model, from \cite{dcc10}}
\end{figure}
In contrast, the
\gls{DCC} has produced a lifecycle model~\cite{dcc10} (\prettyref{f:dcc-lifecycle})
which stresses that data creation, management, and reuse are part of a
cycle in which preservation planning, for example, can naturally
happen before data creation as well as after it; and in which data can
be appraised, reappraised, and possibly disposed of if it becomes
obsolete.  It therefore makes explicit both the short- and long-term
cycles in the flow of active research data, and it emphasizes the
active involvement of data curators in maintaining that cycle.

Cycles of use and re-use are not the only links between datasets.  As
discussed in~\cite{lee07}, one digital object can also provide context
for another, in a variety of ways.  To some extent this remark
rediscovers the notion of the OAIS \gls{Representation Network}, and this
in turn prompts us to stress that although we have contrasted OAIS and
DCC here, they are not in competition: OAIS is concerned with the
creation and management of a working archive with gatekeepers
and firm goals; the DCC model is concerned with the location of the
archive in the wider intellectual context.

The DCC model is immediately compatible with the observation, in
\prettyref{s:preservation-costs} below, that \gls{HEP} and \gls{GW} archives
effectively avoid some preservation costs by seeing long-term
preservation as only part of the role of a data repository.  Accepting
data, making it available as working storage, transforming it into
immediately useful forms, or appraising (possibly regenerable)
datasets whose storage costs outweigh their usefulness, all give the
archive a familiarity with the data, and the researchers a familiarity
with the archive, which means that the decision to select certain data
for long-term preservation is potentially more easily reached, more
easily defended and more easily funded, than if the archive is
conceived as a cost-centre bucket bolted on the side of the project.
This appears to be borne out by the LIGO experience, in which the
new \gls{DMP} plan was developed and promoted by the same personnel who had
long been responsible for the design and management of the data management
system on which everyone's daily work depended.
\index{DCC lifecycle|)}


\section{DMP planning\dash practicalities}
\label{s:planning}

At first glance, the development of a DMP plan appears to be a
burdensome addition to the engineering of a large scientific project.
However, there may not be a huge amount to do in fact.

As we noted above, much large-scale science is in the happy position
of starting off with reasonably functional and adequately resourced
data management systems, simply because the experimental apparatus
will be unusable without them.  That is, the DMP problem \emph{is
  already solved to first order}, and this is corroborated by the
discussion in~\cite[\S3.5]{gray10}, which illustrates that a well-run
big-science project will almost automatically score well on a
benchmarking exercise (\q{AIDA},~\cite{aida09}).  Thus a DMP
planning exercise becomes a
question of formalising and tidying existing practice, in order that
these expensive projects do their duty to society and their funders,
and those funders do their duty to society and to their political
masters.  This is the point of view from which we offer the
following observations.

The sections below are roughly ordered from those with the shortest
time horizons, to those with the longest.  However they are largely
disconnected from each other, and might be better regarded as extended
footnotes to the background of \prettyref{s:technical}.

\subsection{Preservation goals}
\label{s:preservation-goals}

A crucial question, easily skipped, is this: what precisely are the preservation goals?

This question is asked in \prettyref{s:preservation-analysis}, and is
implicit in the discussion in \prettyref{s:preserve-raw}, but one
should not leap to the conclusion that everything should be preserved,
indefinitely, simply because this would be far too expensive.

We have already mentioned the notion of the \gls{Designated Community}.
\begin{itemize}
\item Who are the members of the Designated Community?
\item What are they expected to be able to do with the preserved data?
\item \dots and for how long?
\end{itemize}
There is no generic answer to any of these questions, nor any answer
that is discipline-independent.  As we noted earlier, astronomy data
probably tends to remain scientifically interesting longer than
particle physics data, and may also remain intelligible for longer, so
that for a given quantity of resource, it is reasonable for its target
preservation time to be greater.  This interacts with the observations
in \prettyref{s:storagecosts} about the effects of \q{under-valuation}
of future preserved data, and the apparent conclusion in that section
that if data is preserved beyond some threshold time, it can survive
more-or-less indefinitely.
\begin{takeaways}{planners}
\item It is probably infeasible to preserve all of the collected data,
  and what is preserved will be a function of discipline and
  resources.
\item It is reasonable to throw data away, as long as you do it as
  the conclusion of a deliberate evaluation of the costs and value.
\end{takeaways}
\begin{takeaways}{funders}
\item Funders will have to interact with projects at an early stage in
  order to prioritise preservation goals.
\item The final decision on what to preserve may have to wait until
  costs are clearer, later in the project (see also below).
\end{takeaways}

\subsection{Data release planning}
\label{s:data-release-planning}

When large \glspl{facility} service the work proposals of individual scientists
or small groups, they typically release data by simply making it public in
their facility archive, after an advertised \gls{proprietary period}
during which it is available only to the scientists who requested the
observation or measurement.

Large collaborations\dash in this context meaning HEP collaborations
such as the \gls{LHC} experiments, or \gls{LIGO}, or large-scale
astronomical surveys\dash instead typically (plan to) release data in
large blocks.

The \gls{LIGO} collaboration has agreed an algorithm to release data
when triggered by a range of occurences, including published papers
quoting data, when the collaboration has probed a given volume of
space-time, or when a certain time has elapsed after the start of the
current phase of the experiment; see~\cite{anderson11}, summarised in
\prettyref{s:gw-data-release}, for fuller discussion.  The goal,
during the negotiation with the funder which led up to the agreed
plan, was to balance the collaboration members' need for privileged
access to the data, as a reward for their work in creating the
experiment, with the funder's variously-founded desire to see the data
made public as soon as possible.

The \gls{ATLAS} collaboration is experimenting with a system in which,
rather than release the data, with its numerous attendant
complications, they support a service called
\q{Recast}~\cite{cranmer10}, which will take a phenomenological model
as input from a user, and analyse the data in the light of that model.
This system means that searches can be performed on the data by a
broad class of physicists not directly connected to the collaboration,
without requiring them to become familiar with the detailed structure
of the underlying data.  This is effectively a type of high-level data
product, which lets the collaboration retain control of the data,
without obliging them to document a dataset-based data product (which
might be harder or more expensive than adapting existing analysis
software to form the Recast system), and without exposing them to the
costs of handling external analysis based on misunderstandings of the
data.\footnote{This model of course depends on the data users trusting
  the data producers, and so might be sadly inapplicable to the sort
  of data release which might be demanded of the owners of climate
  data, by data users who seem to believe they are being conspired
  against.} See \prettyref{s:lhc-data-release} for further discussion.

Large astronomical surveys tend to release data either after an
observing season is over, or (more commonly) after each complete pass
over the relevant survey area.  The release is not immediate, but
takes place after data reduction and quality assurance checks.  In
this case, it is usually a higher level data product which is
released.
%

\subsection{Validation}
\label{s:practice-validation}

We discussed the general topic if repository audit in
\prettyref{s:oais-audit}.  There, we described the way in which some
repository audit standards are emerging from the original OAIS work.
Here, we would like to provide a few more practical pointers.

It is possible to imagine several levels of certification, with full
adherence to the ISO standard~\cite{std:ccsds652.1} being the most demanding. One scenario
under discussion by the \gls{TRAC} working group conceives of three levels, labelled
bronze, silver and gold. Bronze would apply to repositories which
obtain certification against the Data Seal of 
Approval\footnote{\longurl{http://www.~datasealof~approval.~org/}}; silver
would be granted to bronze level repositories which in addition
perform a structured self-audit based on the ISO standard; and gold
would be granted to repositories which obtain full external audit and
certification based on the ISO standard.

Test audits of six varied repositories in Europe and the USA were
conducted in the summer of 2011, with a view to trialling the standard
and refining the audit procedure. The results are being written up
within the EU project
APARSEN\footnote{\longurl{http://www.alliance~permanent~access.org/~index.php/~current-projects/~aparsen/}}.
Thus it is expected that in the near future awareness of the new
standard will become widespread, and auditing services will start
to become available.

To achieve certification to the ISO standard, a repository must
satisfy the auditors that it satisfies the metrics defined in the
standard. The aim is not to give a \q{pass/fail} certification, but to
highlight areas for improvement, so the repository might offer or be
expected to have plans for improvement in particular areas.

There are a number of really fundamental requirements that the
repository must meet in order to satisfy the auditors that is can be
considered trustworthy for long-term preservation of its digital
material. These include:
\begin{enumerate}
\item Having a clear mission, preservation strategic plan and
  preservation policies. These terms are defined in the standard but
  in essence refer to the explicit commitment of the organisation to the
  stewardship of the digital objects in its custody, the goals and
  objectives for preservation, and the approach to be taken.
\item Identifying and being aware of the needs of its \gls{Designated
  Community}.
\item Monitoring changes in the external environment that might impact
  the repository's functioning.
\item Identifying risk factors and having succession planning and
  disaster recovery.
\item Making reference to the OAIS information model, particularly
  distinguishing the various \glspl{Information Package} and handling them
  appropriately, and capturing appropriate \gls{Representation
  Information}. OAIS distinguishes between the \gls{SIP} (which is received
  by the repository), the \gls{AIP} (what the repository stores and
  maintains internally), and the \gls{DIP} (given out to accessors of the
  repository). Being aware of these distinctions is important, though
  there is often (or perhaps even usually) significant overlap between
  them, so that the difference is more one of audience than
  significant technical content.
\item Having mechanisms for tracking digital objects through the
  system, and for ensuring their continued integrity.
\end{enumerate}
Even without certification, this list provides a high-level checklist
of planning desiderata.

We believe it would be useful for funders to require basic (for example \q{bronze})
validation of projects, for projects above a certain scale.  A
different level of validation, or none, may be appropriate for projects of a
different scale, or where the funder has different requirements for
the resulting data (for example, one can imagine a funder feeling
obliged to make different curation and visibility requirements for
climate data).  There
is a bureaucratic cost, of course, but this would provide very
straightforward signoff on both sides, and would (we anticipate) be useful
for the design of the project's data management system.  We believe
that most well-run large projects would be able to achieve this
without significant difficulty: as we noted at the beginning of this
section, the data management for a large-scale science project must be
reasonably well-run simply in order for the experiment to function.
Certification would incur some additional costs (as is the case for
ISO-9000 certification, for example); these should be incurred by the funder.

\begin{takeaways}{planners}
\item Even an informal self-audit provides a structured way to unearth
  problems; a self-audit can be used as a type of reassuring
  validation.
\item An unvalidated archive may be of little practical use.
\item Between them, the CASPAR and TRAC outputs provide quite concrete
  advice on implementing an OAIS-inspired plan.
\end{takeaways}
\begin{takeaways}{funders}
\item There are both financial and effort costs associated
  with validating repository designs.
\item There exists emerging good practice for instantiating
  OAIS-inspired designs, but it is not yet stable enough to provide
  check-list requirements (especially since big-science DMP problems
  may always require more or less bespoke solutions).
\item That said, there will soon be concrete validation standards for archives,
  and depending on requirements, it may be useful for both funders and
  projects to refer to these standards in negotiations.
\end{takeaways}

\subsection{Software and service preservation}
\label{s:sw-preservation}
\index{software preservation|(}As discussed above, there is often a
substantial amount of important information encoded in ways which are
only effectively documented in software, or software configuration
information.  There is therefore an obvious case for preserving this
software (though note the caveats of \prettyref{s:preserve-raw}).

Preservation of a software pipeline requires preserving the \gls{pipeline} software
itself, a possibly large collection of libraries the software depends
on, the operating system (OS) it all runs on, and the configuration
and start-up instructions for setting the whole thing in motion.  The
OS may require particular hardware (CPUs or GPUs), the software
may be qualified for a very small range of OSs and library versions,
and it may be hard to gather all of the configuration information
required (there is some discussion of how one approaches this problem
in for example~\cite{south11}).  It is not certain that it is
necessary, however: if the data products are well-enough described,
then re-running the analysis pipeline may be unnecessary, or at least
have a sufficiently small payoff to be not worth the considerable
investment required for the software preservation.  We feel that, of
the two options\dash preserve the software, or document the data products\dash
the latter will generally be both cheaper and more reliable as a way
of carrying the experiment's information content into the future, and
that this tradeoff is more in favour of data preservation as we
consider longer-term preservation.

This last point, about the changing tradeoff, emphasizes that the two
options are not exclusive: one can preserve data \emph{and} preserve
software, and the EPSRC-funded Software Sustainability
Institute\footnote{\url{http://www.software.ac.uk/}} provides a
growing set of resources which provide guidance here.  However the
solutions presented generally focus on active curation, in the sense
of preserving software through continuing use and maintenance (and
thus, as the institute's name suggests, this becomes a question of
sustainability rather than necessarily preservation).  This can be
successful\footnote{The UK Starlink project provided astronomical
  software to the UK and internationally.  It ran from 1980 to 2005, when it was at the last moment
  rescued from oblivion by being taken up by the UK Joint Astronomy
  Centre \ifxetex Hawai‘i\else Hawai`i\fi.  The current distribution includes still-working
  code from the 80s.  The Netlib and BLAS libraries have components
  which date from the~70s.}, and is the approach implicit
in~\cite{south11}; however it means that the sustainability of a piece
of software now depends on the existence and continuing vitality of a
community which can care for it, which means that it is brittle in the
face of significant funding gaps.  This process can be encouraged by a
suitably open process, but while this may possibly need fewer
resources it probably needs more personal commitment, and is even less
predictable than a funded solution.  While it might seem that a
software set without users does not need to be preserved, it might be
unused deliberately, because it is an early software version or
abandoned pipeline strategy which, though later deprecated, is still
necessary to re-generate or validate a historical release of a data
product.  Despite these qualifications, assuming a continued supported
future is still a reasonable preservation strategy, since it
encourages more and better \gls{Representation Information} in the
form of design or user documentation, which can only improve the
software's chances of surviving a support gap.\footnote{We are
  grateful to Rob Baxter of the EPCC for valuable comments on this
  topic.}

The Recast system mentioned in \prettyref{s:data-release-planning}
comes under the heading of software preservation\dash it is software,
and it needs to be preserved.  However it is different from the
preservation targets discussed in this section in that its
preservation is not an afterthought, but instead its preservability
has been designed into it.  This prompts us to at least mention the
problem of \emph{service preservation}.  Preserving services is at once harder and
easier than preserving data.  It is harder, since more infrastructure
has to be present in order for a service to be viable; but easier in
the sense that a service will almost necessarily have useful
Representation Information (or rather its analogue for services rather
than data) in the form of service interface documentation, and it may
be easier to reassure oneself that a service is running, and working
correctly, than it is to reassure oneself that a dataset is actually
intelligible.  The topic of service preservation is not currently
well-understood.\footnote{David Rosenthal makes some interesting
  observations on the challenges of preserving services in
  \url{http://blog.dshr.org/2011/08/moonalice-plays-palo-alto.html}.
  One conclusion from this is that the increasing importance of
  dynamic web content and services means that, when talking
  about the web, the
  distinction between preserving \q{documents} and preserving
  \q{software} is disappearing.}


\index{software preservation|)}

\subsection{Costs and cost models}
\label{s:preservation-costs}

\index{costs|(} 

There is a good deal of detailed information, and some modelling, of
the costs of digital preservation. However this has not turned into a
strong consensus, and it may be that the variation in preservation
contexts means that no simple consensus is possible.  All we can do
here is to highlight some of the work that has been done in this area,
in the hope that this can be used to ground an estimate for a
particular project's preservation costs, in some sort of principle.

Preservation costs can be understood under three broad headings.
\begin{description}
\item[Storage] The most obvious cost of digital preservation is the
  cost of simply preserving the bytes into the future, but this
  ignores the costs associated with getting the data into an archived
  form, and managing its curation.  In the short term this is a
  trivial calculation, and a rather modest cost; but in the \gls{Long
    Term} (in the OAIS sense of more than one technology generation)
  it dominates the cost, and is a complicated function of economic and
  technical assumptions, and preservation goals.  See
  \prettyref{s:storagecosts}.
\item[Ingest and acquisition] Data is not typically generated pre-labelled and ready
  for deposit, and there are significant costs associated with making
  it so ready, involving developing and generating metadata,
  normalising the data, and in some cases sorting out rights-based
  issues.  Depending on what is being archived, ingest costs can
  represent up to 80\% of staff costs, but these costs are
  dramatically reduced if (as is happily often the case for the
  big-science projects this report is nominally addressed to) the data
  is accessed day-to-day in more or less the same form in which it is
  archived.  The design and acquisition costs must still be paid, of
  course, but they are part of a development budget rather than a
  preservation budget, so must only be paid once.  See
  \prettyref{s:ingest} for some more observations on this heading.
\item[Staffing] Ingest may represent a large fraction of a project's
  staff costs, but even separately from that there are costs
  associated with everything from routine system management, to
  supporting experts preserving implicit knowledge by continuing
  active work with the data. There is little more we can usefully say
  about this, beyond remarking that the associated costs will be well
  understood at the local sites where the expenditure happens.
\end{description}

\subsubsection{Existing practice}

There have been a few studies of preservation costs in digital
preservation projects.  These reach some consensus on the main
headings\dash aquisition and ingest is expensive, and costs scale
weakly with archive size\dash but without consensus on an explicit
costs model.  We briefly summarise these below, and then discuss the
some of the differences between these general studies and specifically
science data.  For a few more details on the studies below, see
Sect.~3.4 of~\cite{gray10}.

The KRDS2 study \cite[\S\S6\&7]{beagrie10} includes
detailed costings from a number of running digital preservation
projects, in some cases down to the level of costings
spreadsheets. The LIFE${}^3$ project has also developed predictive
costings tools~\cite{hole10}, and the PLANETS project
(\url{http://www.planets-project.eu/}) has generated a broad range of
materials on preservation planning, including costing studies.

Although there is a broad range of preservation projects surveyed in
the KRDS report, there are numerous common features. Staff costs
dominate hardware costs, and scale only very weakly with archive
size. The study also notes that acquisition and ingest costs are a
substantial fraction (70--80\%) of overall staff costs, but also scale
very weakly with archive size. These are relatively small archives,
generally below a few~\si{TB} in size, where ingest is a significant
component of the workload. In this report we are interested in
archives three or four orders of magnitude larger than this where (as
discussed below) ingest may be cheaper, but in broad terms, it appears
still to be true that (at least in the short term) staff costs dominate hardware costs at larger
scales, and scale only weakly with archive size.

Note that the figures discussed here are (as it turns out) figures for
what one might call \q{live} archives, where the data has an active
user community, which the archive invests resources in supporting,
and in so doing maintains a healthy community of individuals with expertise in
using the data (that is, possessing and sharing the \gls{tacit
  knowledge} of how the data is to be used).  The situation changes somewhat when talking about
long-term preservation, not quite in the OAIS sense
of \gls{Long Term} (which is focused on technology changes), but in
the sense that data is not seen by humans for extended periods, and
where there are, by hypothesis, no walking and talking sources of
advice about the data.  In the case of \q{unaccessed} data, there is even less in the way of
robust cost modelling, although it seems likely that the cost model
for this would be dominated by the costs of byte storage (discussed in
\prettyref{s:storagecosts}) rather than staff costs.

There is probably rather little actual experience of digital archives
working entirely without advice from human curators.  Astronomy archives may come
closest, but this may be atypical, if indeed it is the case that
astronomy data has an in-built tendency to remain intelligible
long-term (as suggested in \prettyref{s:case-data-preservation}).  The
authors of~\cite{curry11}, and in passing~\cite{south11}, describe the
sort of data archaeology which is required in the absence of paper or
personal \gls{Representation Information}.

The lack of scaling with size, even when an archive progressively
grows in size, seems to suggest that it is an archive's \emph{initial}
size (in the sense of small, medium or large, for the time) that
largely governs the costs.

Information from two large astronomy archives~\cite[\S3.4]{gray10} was
found to be consistent.  The two archives held of order
\SI{100}{\tera\byte} each; one spent 25--30 staff-years on initial
development, and both spend in the range of 3--6 staff-years per
year on maintenance and support; each seems to spend between a quarter
and a third of its budget on hardware.  Both archives are funded from
a mixture of short- and long-term grants.

The \gls{NASA} \gls{PDS} has developed a parameterized model for
helping proposers estimate the costs involved in preparing data for
archiving in the
PDS\footnote{\url{http://pds.nasa.gov/tools/cost-analysis-tool.shtml}
  and see~\cite{nasa:ceh08}.};
most relevantly for the above discussion it includes a scaling with
data volume of $1+1.5\log_{10}(\mbox{volume/MB})$.

As noted in \prettyref{s:case-data-preservation}, the \gls{HEP} community is now
constructing more detailed plans for data preservation, and the
associated costs. Reference~\cite{south11} estimates (albeit without
an explicit costs model) that a 
long-term archive would cost 2--3 FTEs for 2--3 years after the end of the
experiment, followed by 0.5--1.0 FTE/year/experiment spent on the
archive's preservation. They compare this to the 100s of FTEs spent on
for the running of the experiment, and on this basis claim an archival
staff investment of 1\% of the peak staff investment, to obtain a
5--10\% increase in output (the latter figure is based on their
estimate that around 5--10\% of the papers resulting from an experiment
appear in the years immediately after the experiment finishes; since
this latter figure is derived on the current model, which achieves
this without any formal preservation mechanisms, this estimate of the
return on investment in archives may be very optimistic).
\begin{takeaways}{planners}
\item There is prior experience of modelling the costs of data
  preservation, with broadly consistent results.
\item These models are not detailed, and are clearly dependent on the
  data type and volume.
\end{takeaways}
\begin{takeaways}{funders}
\item It may be infeasible to make robust estimates of the costs of
  preservation, before a project has gained experience with the final
  form of the gathered data.
\end{takeaways}

\subsubsection{Ingest and acquisition}
\label{s:ingest}

We have repeatedly noted above that in astronomical, HEP and GW contexts, archive
ingest\index{data!ingest} is generally tightly integrated with the
system for day-to-day data management, in the sense that data goes
directly to the archive on acquisition and is retrieved from that
archive by researchers, as part of normal operations. On the other
side of the archive, projects will generate and disseminate data
products\dash which look very much like OAIS \gls{DIP}s\dash as part of
their interaction with external collaborators, without regarding these
as specifically archival objects. Thus the submissions into the
archive may consist of both raw data and things which look very much
like DIPs, and the objects disseminated will include either or both
very raw and highly processed data. The \emph{long-term} planning
represented in the LIGO DMP\index{LIGO!DMP}~\cite{anderson11}, for
example, is therefore less concerned with setting up an archive, than
with the adjustments and formalizations required to make an existing
data-management system robust for the archival long term, and more
accessible to a wider constituency. What this means, in turn, is that
some fraction of the OAIS ingest and dissemination costs (associated
with quality control and metadata, for example) will be covered by
normal operations, with the result that the \emph{marginal} costs of
the additional activity, namely long-term archival ingest and
dissemination, are probably both rather low and typically borne by
infrastructure budgets rather than requiring extra effort from
researchers.\footnote{This is consistent with the ERIM project's
  conclusions that \qq{ideally information management
    interventions should result in a zero net resource
    increase}~\protect\cite[p.8]{darlington11}. In this case
  there is no extra resource required from the researchers, though
  there might be a need for extra resource under an infrastructure
  heading.}

This is corroborated by our informants above, who generally regard
archive costs as coming under a different heading from \q{data
  processing costs}. The point here is not that the OAIS model does
not fit well\dash it fits very well indeed\dash nor that ingest and
dissemination do not have costs, but that if the associated activities
can be contrived to overlap with normal operations, then the costs
directly associated with the archive may be significantly
decreased. This is the intuition behind the recent developments in
\q{archive-ready} or \q{preservation-aware storage} (cf
\cite{factor07} and \prettyref{s:dcc-lifecycle}), and confirms that it
is a viable and effective approach.

As a final point, we note that big-science projects are inevitably
also large-scale engineering projects, so that the consortia and their
funders are inured to the procedures, uncertainties and
management of cost estimates, so that the costing and management of
data preservation can be naturally built in to the relationship
between funders and funded, if the funders so require.

\begin{takeaways}{planners}
\item Despite the prominence of ingest costs in some discussions of
  DMP planning, these may be a relatively minor facet of the cost
  model of large-scale physics projects.
\end{takeaways}

\subsubsection{Modelling storage costs}
\label{s:storagecosts}

While ingest costs may or may not be substantial, they are heavily front-loaded;
and staffing costs, though long-term, are predictable and their
estimation is largely a function of predicted inflation measures.  In
contrast, any estimate of the costs of long-term \emph{storage}\dash the
activity of simply preserving bytes into the future\dash depends
on a broad range of poorly-understood economic variables, and the
necessarily unpredictable effects of future changes in technology.

In a series of blog posts, David Rosenthal has described the ongoing
development of a model for estimating long-term storage
costs~\cite{rosenthal11,rosenthal11a,rosenthal12}.  The model is
purely concerned with storage costs, rather than ingest or
adminstration costs, and takes as its paradigmatic problem the goal of
storing a petabyte for a century.  This is a solved problem, if money
is no object\dash with enough replication, and migration, and
sufficiently rigorously checked checksums, and suitable attention to
novel failure modes, a petabyte can be stored with adequately (though
not arbitrarily) high likelihood of success~\cite{rosenthal10}.

The problem comes in paying for this or, put another way, attempting
to estimate a cost for such preservation which is robust enough that
it is believable, and ideally low enough not to cause the preservation
community to throw up its hands in despair and think longingly of clay
tablets.

The discussion focuses on \q{Kryder's Law}, which is the observation
that the cost of disk space has been decreasing roughly exponentially
for about three decades~\cite{smith12}.  It is not clear that this
decrease will continue indefinitely into the future, or with the same
power, so that a storage 
model which assumes that it will, implicitly or explicitly, may be in
trouble.

Rosenthal discusses three business models for long-term storage:
(i)~an \q{S3 model}, where a storage provider simply charges rent for
storage, and can increase this rent if the price of storage increases
for some reason (this is not vulnerable to deviations from Kryder's
law, in the sense that a change in Kryder's law will result in a
quantitative rather than qualitative change to the model from the
user's point of view); (ii)~a \q{Gmail model}, where a provider funds storage from
adverts, and hopes that the increase in required storage is balanced
by a greater-than-proportional Kryder's law decrease in the per-GB
cost; and (iii)~an \q{endowment model}, where a quantity of data is
deposited along with a financial endowment to cover the costs of its
preservation into the indefinite future.  Discounting the first two
options as too vulnerable to external factors to be viable
archival strategies, the third option transforms into the question of
how much, per~TB, this initial endowment should be.

Space, power and cooling account for around 60\% of the three-year
cost of a server, and other estimates suggest that media accounts for
between a third and a quarter of the total cost of storage.  Combining
these figures with some rather simple assumptions about the future,
Rosenthal suggests that a markup of two to four times the initial storage cost
(depending on assumptions) will preserve the data reliably, and notes
that Princeton have gone for the lower end of this range and are
charging their own researchers \$3\,000/TB for long-term
preservation~\cite{rosenthal11}.  He concludes that:
\begin{quotation}
Endowing data has some significant advantages over the competing
business models when applied to long-term data preservation. But the
assumptions behind the simple analysis are optimistic. Real endowed
data services, such as Princeton's, need to charge a massive markup
over the cost of the raw storage to insulate themselves from this
optimism. The perceived mismatch this causes between cost and value
may make the endowed data model hard to sell.~\cite{rosenthal11}
\end{quotation}

Subsequent posts in this series discuss the appropriate model for
discounting future cash-flows, the unexpectedly
large effects of even a mild (5\range 10\%) under-valuation of the
preserved data~\cite{rosenthal11a,haldane11}, and the still unsettled
nature of the relationship between the costs of local and cloud storage~\cite{rosenthal12a}.
The work is concerned with the development of a
Monte Carlo model of the preservation process, incorporating long-term
economic yields, the effects of hypothetical new technologies, and
various scenarios for the future of Kryder's law.  The results are as
yet inconclusive, but suggest that endowment multipliers of 4\range 6
are required, and appear to suggest a robust effect where the
probability that a dataset will survive for 100 years, without running
out of money, changes from near~0 to near~1 over a remarkably
small range of around 0.5 in the multiplier~\cite{rosenthal12}.  Also,
this modelling reveals that as the
Kryder's law annual decrease heads down into the 10\range20\% range,
this bankruptcy probability (or specifically, the location of this
threshold) becomes increasingly unpredictable, in the sense of being
increasingly sensitive to model assumptions.  The Kryder's law
decrease is indeed currently heading into this unstable range.

This analysis appears to suggest petabyte-for-a-century endowment costs approaching \$30\,000/TB.

\begin{takeaways}{funders}
\item There is considerable uncertainty in the costs of data storage
  beyond about a decade.
\item What appears to be the best-justified long-term preservation
  model appears to require a large up-front payment in the form of an
  endowment.
\end{takeaways}

\index{costs|)}

\subsection{Modelling data loss}
\label{s:storage}

Quite apart from the difficult problem of modelling the cost of storage, which
includes the cost of hedging against data loss, the underlying
processes of data loss are still imperfectly understood.

Baker et al.~\cite{baker06} discuss a variety of modes for data loss,
along with listing some tempting but dangerous assumptions, and develop a simple
probabilistic model for data loss, which concentrates on the interplay
between \q{visible} faults (by which they mean detected data errors)
and \q{latent} ones (where data has been corrupted or lost, but not
yet detected).  This allows them to examine trends in irrecoverable
data loss rates in a range of replication and checking scenarios.
Though this allows the authors to be quite precise in teasing out how
different aspects of preservation strategies have their effect on loss
rates, which of course has implications for the cost-effectiveness of
those strategies, they remain properly cautious about the detailed
predictive power of their model, and instead confine themselves to
identifying the extent to which different strategies trade off against
each other, and which strategies have the biggest effect on reducing
rates of irrecoverable data loss.

Several of the strategies depend on one or another form of
\emph{replication}, and this strategy is taken to one extreme in the
LOCKSS system\footnote{\url{http://www.lockss.org}}, which is
concerned with preserving library access to journal articles.  The
LOCKSS system depends on libraries preserving separate copies of
articles, in a loosely-coordinated way which allows them to cooperate
to repair detected damage to each other's holdings.  Though this
system is concerned with article data rather than science data, and is
at a somewhat smaller scale than is of immediate concern to the \q{big
  science} readers of this report, it illustrates one extreme of a
replication strategy: data is preserved with rather high
assurance, not as the result of anything technically exotic or
particularly expensive, but instead by stressing independence and
heterogeneity, and that \q{lots of copies keep stuff safe}.

\appendix

\section{Case studies in preservation}

\subsection{ISIS}
\label{s:isis-case}

\subsubsection{Introduction to ISIS}

ISIS is one of major \glspl{facility} operated by STFC at the Rutherford
Appleton Laboratory.  ISIS is the world's leading pulsed spallation
neutron source. It runs 700 experiments per year performed by 1,600
users on the 22 instruments that are arranged on the beamlines. These
experiments generate \SI1{TB} of data in 700,000 files. All data ever
measured at ISIS over twenty years is stored, some 2.2 million files
in all. ISIS is predominantly used by UK researchers, but includes
most European countries through bilateral agreements and EU-funded
access. There are nearly 10,000 people registered on the ISIS user
database. The user base is expanding significantly with the arrival of
the Second Target Station.

\subsubsection{ISIS data}

On ISIS today, the instrument computers are closely coupled to data acquisition electronics and the main neutron beam control. Data is produced in two formats: the ISIS-specific RAW format and the more widespread NeXus format. Access is at the instrument level indexed by experiment run numbers. Beyond this data management comprises a series of discrete steps. RAW files are copied to intermediate and long-term data stores for preservation. Reduction of RAW files, analysis of intermediate data and generation of data for publication is largely decoupled from the handling of the RAW data. Some connections in the chain between experiment and publication are not currently preserved. DOIs are issued for datasets at the experiment level. At present all data is retained.

The data is kept for the long term in archival store: a layered system with three local checksummed copies on mirrored spinning disk, a tape backup and as a dark archive.

Future data management will focus on development of loosely coupled components with standardised interfaces allowing more flexible interactions between components. The ICAT metadata catalogue sits at the heart of this new strategy. It systematically catalogues data files and implements policy controlling access to files and metadata and uses single authentication to allow linking of data from beamline counts through to publications and to support search across facilities.

\subsubsection{The ISIS data policy}

The ISIS data policy~\cite{isis-policy} establishes an understanding of responsibilities and rights of data producers and user,s and of the ISIS \gls{facility} itself.

The policy is structured as follows.
\begin{description}
\item[1. General principles]
These define the scope of the policy and make it clear that adherence is manadtory for ISIS users.
\item[2. Definitions]
Raw data is distinguished from results (\qq{intellectual property, and outcomes arising from the analysis of raw data}), while metadata is defined as \qq{information pertaining to data collected from experiments performed on ISIS instruments, including (but not limited to) the context of the experiment, the experimental team (in accordance with the Data Protection Act), experimental conditions and other logistical information.}
\item[3. Raw data and associated metadata]
Raw data and metadata that is obtained from free (non-commercial) use of ISIS is declared to be in the public domain with ISIS acting as custodian. There is a commitment to curate data for the long term. Data will become publicly accessible after a three-year embargo period, though registration will always be required for access.  The catalogue will link data to proposals, but access to the proposals themselves will not be public.
\item[4. Results]
Ownership of results (as defined above) is determined by the contractual conditions pertaining to the work. ISIS undertakes to store results that are uploaded, but not to fully curate them. Access to results is restricted to those who performed the analysis.
\item[5. Good practice for metadata capture and results storage]
This section encourages provision of good quality metadata and of suitable cooperation and acknolwdgement if data is to be reused by others.
\item[6. Publication information]
It is required that references to publications related to experiments
carried out at ISIS must be deposited in the STFC e-Pubs system
(institutional repository) within six months of the publication date.
\end{description}

\subsection{LIGO/GEO/Gravitational Waves}
\label{s:ligo-case}
The gravitational wave community has astronomical goals, but in the
scale of the LIGO project, and in the amount of novel technology
involved, as well as in the fact that many of the personnel involved
came originally from a HEP background, the project's culture more
closely resembles that of a HEP experiment than of an astronomical
telescope.

\subsubsection{Gravitational wave consortia}
There are three principal sources of recent GW data available to UK
researchers: \gls{LIGO}, \gls{GEO600} and \gls{Virgo}. There are other detectors which
are either smaller efforts (in terms of consortium sizes), which have
stopped taking data (TAMA-300), or which are still at the planning
stage. See~\cite{pitkin11} for an overview of current detectors, and
of detector physics.

While LIGO is a detector, the scientific collaboration which uses it
is known as the \gls{LSC}, which is a network of \glspl{MOU}
between LIGO Lab and other institutions of various sizes.  
In total (as of June 2010), the LSC consists of a
little over 1300 \q{members}; of these, 615 spend more than 50\% of
their time dedicated to the project and so have a place on the LSC
author list.

The Italian/French \gls{Virgo} consortium has its own detector and
analysis \gls{pipeline}, and has a data-sharing agreement with the
LSC, represented by the \gls{LVC}.  Virgo has 246 members (with a
slightly different definition from the LSC), and GEO600 around 100.

Both the LIGO and Virgo detectors will shut down from late-2011 until
roughly 2015, when they will restart with enhanced sensitivity.

\subsubsection{GW data}
Although the consortia have (as expected) announced no detection so
far, they nonetheless produce a large volume of auxiliary data,
representing background and calibration signals of various types, and
this, together with the core data, means that the LSC collectively
produces data at a rate of approximately one \si{\PBY}.

We can readily identify multiple levels of data.
\begin{description}
\item[Raw data] The lowest-level GW data consists of the signals from
  the core detectors.\index{raw data} This data is made meaningful
  only by processing with software which is completely specific to the
  detectors in question. This is stored in \q{frame format}, which is a
  very simple format intelligible to all the primary data analysis
  software in the community, and which is multiply replicated across
  North America, Europe and Australia. Although the disk format is
  common, the semantic content of the raw data is specific to
  detectors and software, so that preserving it long-term would
  represent a significant curation challenge. 
\item[Data products] The raw data is processed into calibrated
  \q{\gls{strain data}}, which is the data channel in which a GW
  signal will eventually be found (this is possibly, but not
  necessarily, also held in frame format). This is the class
  of data products\index{data products} which will eventually be made
  public. Unusually, it turns out that GW raw data is in a
  semi-standard format, and the data products are specific to the
  analysis \gls{pipeline} which produced them.
\item[Publications] Sitting above the data products is a class of high-level data products, scientific papers, and other peer-reviewed outputs. The GW projects have announced no detections of gravitational waves, but have nonetheless produced a broad range of astrophysically significant negative results~\cite[\S6.2]{pitkin11}.
\end{description}

Both the \q{data product} and \q{publication} groups are broad classes
of objects. The practical boundary between them is clear, however:
what we are calling \q{publications} are entities such as journal
articles or derived catalogues whose long-term curation is not the
responsibility of the LSC data archive, though they may be held in
some separate LSC paper archive.

\subsubsection{Gravitational wave data release}
\label{s:gw-data-release}
Because the \gls{LSC} has not announced the detection of any signal so far,
and because the data will remain proprietary to the consortium until
well after such an announcement\glsadd{proprietary period}, there are no
distributed data products so far, and so the issues surrounding
formats and documentation have not yet been addressed. However it is
the eventual public data products which are the highest-value outputs
from the experiment, and which are the products which it will be most
important to archive indefinitely.

At present, \gls{LIGO} data is available only to members of the
\gls{LSC}. This is an open collaboration, and research groups which
join the LSC have access to all of the LIGO
data\footnote{\longurl{http://www.ligo.org/~about/join.php}}. In return,
they contribute personnel to the project (including for example people
to do shift-work manning the detectors), and accept the
collaboration's publication policies, which require that all
publications based on LIGO data are reviewed by the entire
collaboration, and carry the complete 800-person author list. At
present, and in the future, data which is referred to by an LSC
publication is made publicly available. 

The LIGO collaboration's future plans for data curation and release
are described in the collaboration's exemplary DMP
plan~\cite{anderson11}.

The LIGO plan proposes a two-phase data release scheme, to come
into play when \gls{aLIGO} is commissioned; this was prepared at the
request of the \gls{NSF}, developed during 2010\range11, and will be
reviewed yearly.

The plan documents the way in which the consortium will make \gls{LIGO} data
open to the broader research community, rather than (as at present)
only those who are members of the \gls{LSC}.  This document describes the
plans for the data release and its proprietary periods, and outlines
the design, function, scope and estimated costs\index{costs} of the
eventual LIGO archive, as an instance of an \gls{OAIS} model.  This is a
high-level plan, with much of the detailed implementation planning
delegated to partner institutions in the medium term.

In the first phase, data is released much as it is at present:
validated data will be released when it is associated with detections,
or when it is related to papers announcing \emph{non}-detections (for
example, associated with another astronomical event which might be
expected or hoped to produce detectable GWs).  In the second phase\dash
after detections have become routine, and the LIGO equipment is acting
as an observatory rather than a physics experiment\dash the data will be
routinely released in full: \qq{the entire body of gravitational wave
  data, corrected for instrumental idiosyncrasies and environmental
  perturbations, will be released to the broader research
  community. In addition, LIGO will begin to release near-real-time
  alerts to interested observatories as soon as LIGO \emph{may} have
  detected a signal}~\cite[\S1.2.2]{anderson11}.  This second phase will begin after LIGO has
probed a given volume of space-time (see \cite[ref 7]{anderson11}),
\emph{or} after 3.5 years have elapsed since the formal LIGO
commissioning, whichever is earlier.  Alternatively, LIGO may elect
to start phase two sooner, if the detection rate is higher than
expected.

In phase two, the data will have a 24-month \gls{proprietary period}.

The DMP describes three (OAIS) \glslink{Designated Community}{Designated Communities}.
Quoting from~\cite[\S1.5]{anderson11}, the communities are as follows.
\begin{itemize}
\item LSC scientists: who are assumed to understand, or be responsible for, all the complex details of the LIGO data stream.
\item External scientists: who are expected to understand general
  concepts, such as space-time coordinates, Fourier transforms and
  time-frequency plots, and have knowledge of programming and
  scientific data analysis. Many of these will be astronomers, but
  also include, for example, those interested in LIGO's environmental
  monitoring data. 
\item General public: the archive targeted to the general public,
will require minimal science knowledge and little more computational
expertise than how to use a web browser. We will also recommend or
build tools to read LIGO data files into other applications.
\end{itemize}

The LIGO DMP plan is, we believe, a good example of a plan for a project of
LIGO's size: it is specific where necessary, it was negotiated with the
project's funder (\gls{NSF}) so that it achieved their goals, and it went
through enough iterations with the broader LIGO community (the agreed
version in \cite{anderson11} is version~14) that its authors could be
confident it had their approval, and that the community was
comfortable with what the DMP plan was proposing.  The document has a
strong focus on the \gls{LIGO} data release criteria, since this was
the most immediate concern of both the funder and the project, but it
systematically lays out a high-level framework for future data preservation,
guided by the \gls{OAIS} functional model.

\subsection{LHC experiments}
\label{s:lhc-data-release}
There is as yet no agreed general policy on data openness and
curation for the \gls{LHC} experiments, but an active discussion is
underway. \gls{CMS} has approved a trial policy, while others are
still evaluating the options.

The investment in LHC data is at a level that requires effort be made
to consider how it might be made available for future use. A set of
communities that would use this facility is easily identified.
\begin{itemize}
\item Original collaboration members long after data taking.
\item The wider \gls{HEP} and related communities
\item Those in education and outreach.
\item Members of the public with an interest in science.
\end{itemize}
One possible response that would require immediate and additional
ongoing resources is for LHC experiment data to be open access after a
period of a few years; this is the basis of the CMS trial.

Another approach would be to retain the data and analysis environment
in-house and allow analysis by people inside and outside of the
collaboration though a well-defined interface. This is the basis of
the Recast~\cite{cranmer10} system, currently finding favour in \gls{ATLAS}.

The first approach has the advantage of full openness and the larger
potential for extending the analyses, but is resource-hungry and
assumes the capture of a great deal of \gls{tacit knowledge}. The second
approach has advantages in terms of support costs and is likely to
encourage robust results. 

Different users will require different levels of data abstraction. Four levels of abstraction emerge.
\begin{description}
\item[Level 1] Supporting documents and any additional numerical data,
  to be released concurrently with the publication and made available
  in public sources such as open access journals,
  INSPIRE\footnote{\url{http://projecthepinspire.net/}} or
  HEPData\footnote{\url{http://hepdata.cedar.ac.uk/}}.
\item[Level 2]  Simplified high level data formats that allow for
  simple reanalysis. This could be for theory comparison, or simply
  education and outreach. 
\item[Level 3] The full analysis data chain post-reconstruction. This
  would allow serious reanalysis but would require the latest analysis
  software and calibrations available through the same computer
  systems that hold the archived data. Only a subset of the available
  integrated luminosity would be made open while there was a prospect
  of increasing the sample.
\item[Level 4] This is the full raw offline data and the software
  necessary to redo reconstruction together with the necessary
  documentation. The software would have to be freely available under
  license.  Only a subset of the data need be available while the
  experiment is still taking data. Continuing access to the full
  databases would be required for use of level~4 data. These data
  would need to be covered by a Creative Commons waiver with an
  associated \gls{DOI} for citation purposes
\end{description}

There seems to be an emerging consensus that the costs and potential
benefits do not warrant making the Level~4 data generally
available. All experiments already make Level~1 data available through
established mechanisms. The Recast mechanism effectively grants access
to Level~1 and most of Level~2 data. The CMS trial will make the first
three levels available, though with a fixed processing version. 

An alternative to making the level~4 data generally available would be
to provide experiment-hosted services that enable extensions to
analyses that require rerunning reconstruction and simulation
software. This approach would mean that essentially the reanalysis
would be done using the normal data and software channels. This would
be simpler and probably lead to fewer mistakes.

Whatever the technical solution chosen by a given collaboration,
issues concerning the membership of the large collaborations
emerge. The principle incentive to build and operate the experiments
is access to the data and a shared understanding of that data, and the
right to sign subsequent publications. Collaborations may wish to
consider the imposition of conditions such as the following on the use
of public data:
\begin{enumerate}
\item Whenever data is reused, the collaboration that collected it and
  LHC accelerator team must be cited.
\item While avoiding any right of veto of external use, any member of
  the collaboration at the time of publication should have the right
  of authorship on all such papers.
\end{enumerate}

\section{STFC Data principles}
\label{s:stfc-principles}
For convenience, we reproduce the STFC data principles here. For the
original versions, plus STFC's \q{recommendations for good practice},
see~\cite{stfc-data-policy}.  We discuss the
relationship between these and the RCUK principles in
\prettyref{s:rcuk-principles} above.

\subsection{General principles}

SP1. STFC policy incorporates the joint RCUK principles on data management and sharing.

SP2. Both policy and practice must be consistent with relevant UK and international legislation. 

SP3. For the purposes of this policy, the term \q{data} refers to (a) \q{raw} scientific data directly arising as a result of experiment/measurement/observation; (b) \q{derived} data which has been subject to some form of standard or automated data reduction procedure, e.g.\ to reduce the data volume or to transform to a physically meaningful coordinate system; (c) \q{published} data, i.e. that data which is displayed or otherwise referred to in a publication and based on which the scientific conclusions are derived. 

SP4. STFC is not responsible for the use made of data, except that made by its own employees. 

SP5. Data management plans should exist for all data within the scope of the policy. These should be prepared in consultation with relevant stakeholders and should aim to streamline activities utilising existing skills and capabilities, in particular for smaller projects. 

SP6. Proposals for grant funding, for those projects which result in the production or collection of scientific data, should include a data management plan. This should be considered and approved within the normal assessment procedure. 

SP7. Each STFC operated facility should have an ongoing data management plan. This should be approved by the relevant facility board and, as far as possible, be consistent with the data management plans of the other facilities. 

SP8. Where STFC is a subscribing partner to an external organisation, e.g.\ as a member of CERN, STFC will seek to ensure that the organisation has a data management policy and that it is compatible with the STFC policy. 

SP9. Data management plans should follow relevant national and international recommendations for best practice. 

SP10. Data resulting from publicly funded research should be made publicly available after a limited period, unless there are specific reasons (e.g.\ legislation, ethical, privacy, security) why this should not happen. The length of any \gls{proprietary period} should be specified in the data management plan and justified, for example, by the reasonable needs of the research team to have a first opportunity to exploit the results of their research, including any IP arising. Where there are accepted norms within a scientific field or for a specific archive (e.g.\ the one year norm of ESO) they should generally be followed. 

SP11. \q{Published} data should generally be made available within six months of the date of the relevant publication. 

SP12. \q{Publicly available} means available to anyone. However, there may a requirement for registration to enable tracking of data use and to provide notification of terms and conditions of use where they apply. 

SP13. STFC will seek to ensure the integrity of any data and related metadata that it manages. Any deliberate attempt to compromise that integrity, e.g.\ by the modification of data or the provision of incorrect metadata, will be considered as a serious breach of this policy.

\subsection{Recommendations for good practice}

SR1. STFC recommends that data management plans be formulated following the guidance provided by the Digital Curation Centre.\footnote{\url{http://www.dcc.ac.uk/resources/data-management-plans}} STFC (e-Science department) can provide advice upon request.

SR2. STFC would normally expect data to be managed through an institutional repository, e.g.\ as operated by a research organisation (such as STFC), a university, a laboratory or an independently managed subject specific database. The repository(ies) should be chosen so as to maximise the scientific value obtained from aggregation of related data. It may be appropriate to use different repositories for data from different stages of a study, e.g.\ raw data from a crystallographic study might be deposited in a \gls{facility} repository while the resulting published crystal structure might be deposited in an International Union of Crystallography database.

SR3. Plans should provide suitable quality assurance concerning the extent to which data can be or have been modified. Where \q{raw} data are not to be retained, the processes for obtaining \q{derived} data should be specified and conform to the standard accepted procedures within the scientific field at that time.

SR4. Plans may reference the general policy(ies) for the chosen repository(ies) and only include further details related to the specific project. It is the responsibility of the person preparing the data management plan to ensure that the repository policy is appropriate. Where data are not to be managed through an established repository, the data management plan will need to be more extensive and to provide reassurance on the likely stability and longevity of any repository proposed.

SR5. Plans should cover all data expected to be produced as a result of a project or activity, from \q{raw} to \q{published}.

SR6. Plans should specify which data are to be deposited in a repository, where and for how long, with appropriate justification. The good practice criteria assume that this data is accompanied by sufficient metadata to enable reuse. It is recognised that a balance may be required between the cost of data curation (e.g.\ for very large data sets) and the potential long term value of that data. Wherever possible STFC would expect the original data (i.e. from which other related data can in principle be derived) to be retained for the longest possible period, with ten years after the end of the project being a reasonable minimum. For data that by their nature cannot be re-measured (e.g.\ earth observations), effort should be made to retain them \q{in perpetuity}.

\section*{Acknowledgements}
\label{s:acknowledgements}
\addcontentsline{toc}{section}{Acknowledgements}
We are most grateful to the various people who provided comments on
earlier drafts of this document:
Rob Baxter (EPCC),
Peter Clarke (Edinburgh and STFC Computing Advisory Panel),
Catherine Jones (STFC)
and
David Shotton (Oxford Zoology).
We would also like to thank Simon Hodson, programme manager of JISC's 
Research Data Management programme, for supporting this work.

\section*{About the authors}
\addcontentsline{toc}{section}{About the authors}

Juan Bicarregui (STFC) is division head within the e-Science Centre at
STFC, with responsibility for data and information services and
development.  He is project lead of the European PaNData project which
is developing cross-European infrastructure for large-scale national
photon and neutron sources.  He is Chair of RCUK Research Outputs
Network Subgroup on Research Data which recently published Common
Principles on Data Policy.

Norman Gray (Physics and Astronomy, Glasgow)
has a background in particle theory, solar physics, and
microlensing data analysis, but for the last decade or so has been
principally involved in astronomical software development (as part of
the UK Starlink project), the emerging virtual observatory (as part of
the EuroVO and Astrogrid projects, while based at the universities of
Glasgow and Leicester), the intersection of the semantic web and
astronomy, and the problems of large-scale science data management.
He is currently the chair of the IVOA's Semantics Working Group,
and has been the co-author of a number of IVOA standards.

Robert Henderson (Physics, Lancaster) is a physicist-programmer with nearly
40 years experience in fixed target, ep-collider and now pp-collider
experiments. He is expert in the event and analysis data formats and
reconstruction and analysis lifecycle of thes eexperiments, and is
currently the ATLAS software release co-ordinator. Having a strong
background in the software, computing and also real physics analysis,
he is well placed to devise realistic data management plans, including
all of the real world practical pitfalls.

Roger Jones (Physics, Lancaster)
is the ATLAS-UK Computing Co-ordinator,
ATLAS Computing Upgrade task leader and previously chaired the ATLAS
International Computing Board. He was on the panel that drafted the
WLCG MOU. These roles make him deeply aware of the political and
multi-national aspects of data management and preservation. His long
physics experience means he is familiar with the issues and policies
attempted by the previous two generations of experiments from the
1980s and 1990s. He also led the ATLAS component of ESLEA project,
exploiting switched optical lightpath technologies for particle
physics, which entailed work on the end storage systems. He was GridPP
Applications Co-ordinator and is a member of the GridPP Project
Management Board. He has had a long-term responsibility for the ATLAS
computing model and is part of the small team assessing the ATLAS data
dissemination, archival and long-term curation. He is Director of High
End Computing at Lancaster University.

Simon Lambert (STFC) is a Project Manager in the e-Science Centre at
STFC, currently specialising in digital preservation. He is the STFC
manager for SCAPE, a large-scale European-funded project developing
scalable solutions for digital preservation in a variety of domains
including scientific research data. He also participated in the
PARSE.Insight project, which produced a roadmap for the future
European e-infrastructure in preservation. He is an active member of
the group working towards an ISO standard for audit and certification
of digital repositories.

Brian Matthews (STFC) is leader of the Scientific Information Group in
the e-Science Centre at STFC. He led the development of the ICAT
metadata model which forms the basis for data management at ISIS and
other large facilities. He led the STFC component of the JISC project
I2S2 (Infrastructure for Integration in Structural Sciences), which
worked towards a data-driven research infrastructure across the
structural sciences.Moreover, his group has been involved in several
projects in digital preservation; as part of the CASPAR project an
approach to preservation objectives and network modelling was
developed.

\section*{Document history}
\addcontentsline{toc}{section}{Document history}
\label{s:documenthistory}

\begin{description}
\item[v0.1, 2012 March 14] First public release of draft
\item[v0.2, 2012 May 7] Responses to initial comments.
\item[v1.0, 2012 August 17] Version 1.0
\end{description}

\clearpage
\addcontentsline{toc}{section}{Glossary}
\def\glossarypreamble{\emph{Terms marked \q{OAIS} are copied from the
    OAIS specification~\cite[\S1.7.2]{std:oais}. Readers of this
    document might also be interested in the Research Data Management
    glossary maintained at \url{http://vocab.bris.ac.uk/data/glossary/}}}
\def\glsgroupskip{} 
\def\delimR{--}   
\printglossaries

\clearpage
\bibliography{mardi-gross}
\bibliographystyle{unsrturl}

\end{document}